\documentclass[12pt,english]{article}
\usepackage{geometry}
\usepackage{amsmath}
\usepackage{amssymb}
\usepackage{stackrel}

\usepackage[english]{babel}
\usepackage[T1]{fontenc}
\usepackage{lmodern}
\usepackage[utf8]{inputenc}
\usepackage{mathrsfs}
\usepackage{float}
\usepackage[numbers,sort&compress]{natbib}

\usepackage{url}
\usepackage{hyperref}
\hypersetup{
  colorlinks   = true, %colors links instead of ugly boxes
  urlcolor     = blue, %color for external hyperlinks
  linkcolor    = blue, %color of internal links
  citecolor   = blue %color of citations
}
\allowdisplaybreaks
\usepackage{breqn}
\usepackage{enumitem}
\usepackage{graphicx}
\usepackage{xspace}

\makeatletter
\let\oldmaketitle\maketitle% Store old \maketitle
\renewcommand{\maketitle}{% Update \maketitle
\begin{singlespace}
  \oldmaketitle% Use singlespace
\end{singlespace}}

\newcommand{\lsim}{\;\raisebox{-.3em}{$\stackrel{\displaystyle <}{\sim}$}\;}

\newcommand\citeres[1]{Refs.~\cite{#1}}
\newcommand\refeq[1]{Eq.~(\ref{#1})}
\newcommand\refeqs[1]{Eqs.~(\ref{#1})}
\newcommand\refse[1]{section~\ref{#1}}

\def\reffi#1{\mbox{Figure~\ref{#1}}}
\newcommand{\gev}{\,\, \mathrm{GeV}\xspace}
\newcommand{\tev}{\,\, \mathrm{TeV}\xspace}
\newcommand\LP{\left(}
\newcommand\RP{\right)}

\def\order#1{\ensuremath{{\cal O}(#1)}}
\newcommand{\Ztp}{\ensuremath{\mathbb{Z}'_2}\xspace}

\usepackage{soul}
\usepackage{xcolor}
\definecolor{Orange}{named}{orange}
\definecolor{Purple}{named}{purple}
\definecolor{Lightblue}{cmyk}{0.9,0.1,0.1,0.3}
\definecolor{dgelborange}{cmyk}{0.,0.3,0.5, 0.}
\definecolor{Lila}{rgb}{0.5,0.,1}
\definecolor{Darkgreen}{rgb}{0.,.7,0.2}
\makeatother

\begin{document}

\sloppy

\thispagestyle{empty}

\def\thefootnote{\fnsymbol{footnote}}

\begin{flushright}
  DESY-21-034~~\\
  IFT-UAM/CSIC-21-018~~
\end{flushright}

\begin{center}
{\Large \textbf{Fate of electroweak symmetry in the early Universe:\\
[0.4em] Non-restoration and trapped vacua %, phase transitions [0.4em] %\htb{and beyond}\htr{??} 
in the N2HDM}}  

\vspace{0.8cm}

Thomas~Biekötter$^{1}$\footnote{thomas.biekoetter@desy.de},
Sven~Heinemeyer$^{2,3,4}$\footnote{Sven.Heinemeyer@cern.ch},
Jos\'e~Miguel~No$^{2,5}$\footnote{josemiguel.no@uam.es},\\[0.4em]
Mar\'ia~Olalla~Olea$^{1}$\footnote{maria.olalla.olea.romacho@desy.de}
and~Georg~Weiglein$^{1,6}$\footnote{georg.weiglein@desy.de}

\vspace{0.8cm}

\textsl{$^{1}$Deutsches Elektronen-Synchrotron DESY, Notkestraße
85, D-22607 Hamburg, Germany}

\vspace{0.1cm}

\textsl{$^{2}$Instituto de Física Teórica UAM-CSIC, Cantoblanco,
28049, Madrid, Spain}

\vspace{0.1cm}

\textsl{$^{3}$Campus of International Excellence UAM+CSIC, Cantoblanco,
28049, Madrid, Spain}

\vspace{0.1cm}

\textsl{$^{4}$Instituto de Física de Cantabria (CSIC-UC), 39005,
Santander, Spain}

\vspace{0.1cm}

\textsl{$^{5}$Departamento de F{í}sica Te{ó}rica, Universidad
Aut{ó}noma de Madrid (UAM), }\\
\textsl{ Campus de Cantoblanco, 28049 Madrid, Spain}

\vspace{0.1cm}

\textsl{$^{6}$II.\  Institut f\"ur  Theoretische  Physik, 
Universit\"at  Hamburg, Luruper Chaussee 149, }\\
\textsl{D-22761 Hamburg, Germany}

\begin{abstract}
Extensions of the Higgs sector of the Standard Model allow for a rich cosmological history around the electroweak scale. 
We show that besides the possibility of strong first-order phase transitions, which have been thoroughly studied in the literature, 
also other important phenomena can occur, 
like the non-restoration of the electroweak symmetry or the  existence of vacua in which the Universe becomes trapped,
preventing a transition to the electroweak minimum.
Focusing on the next-to-minimal two-Higgs-doublet model (N2HDM) of type~II and taking into account
the existing
theoretical and experimental constraints, we identify the scenarios
of electroweak symmetry non-restoration, vacuum trapping and first-order 
phase transition
in the thermal history of the Universe. We analyze these phenomena and in particular their relation to each other, and discuss their connection
to the predicted phenomenology of the N2HDM at the LHC. 
Our analysis demonstrates that the presence of a global electroweak minimum of the scalar potential at zero temperature does not guarantee that the corresponding N2HDM parameter space will be physically viable: 
the existence of a critical temperature at which the electroweak phase becomes the deepest minimum is not sufficient for a transition to take place, 
necessitating an analysis of the tunnelling probability to the electroweak minimum for a reliable prediction of
the thermal history of the Universe. 

\end{abstract} 

\end{center}

\newpage

\renewcommand{\thefootnote}{\arabic{footnote}}
\setcounter{footnote}{0}

\clearpage
\setcounter{page}{1}

\tableofcontents

\section{Introduction}
\label{sec:intro}
The discovery of a Higgs boson with a mass 
of about 125~GeV at the Large Hadron Collider (LHC)~\cite{Aad:2012tfa,Chatrchyan:2012ufa} 
was a milestone in our understanding of the laws of nature. 
Within the current experimental and theoretical uncertainties, 
the properties of the detected particle agree with the predictions of the Standard Model (SM)~\cite{Khachatryan:2016vau,Aad:2019mbh,Sirunyan:2018koj}.
However, they are also compatible with a wide variety of extensions of the SM that are motivated in view of 
several shortcomings of the SM which require new physics beyond the SM (BSM). Among others, the ingredients of the SM are not sufficient to generate the observed matter-antimatter asymmetry of the Universe~\cite{Huet:1994jb,Gavela:1994ds,Gavela:1994dt}, and the SM lacks a particle candidate to explain the observed cosmological abundance of dark matter~\cite{Aghanim:2018eyx}. 

Extensions of the SM scalar sector, e.g.~by gauge singlets or $SU(2)$ doublets, provide 
scope to address the above shortcomings. 
For instance, adding further Higgs doublets to the SM~\cite{Lee:1973iz,Kim:1979if,Wilczek:1977pj} allows the generation of the observed matter-antimatter of the Universe via electroweak (EW) baryogenesis~\cite{Cline:1996mga,Fromme:2006cm,Cline:2011mm,Dorsch:2016nrg} (see~\cite{Cohen:1993nk,Trodden:1998ym,Morrissey:2012db} for general reviews on EW baryogenesis). %
A necessary ingredient in such a case is a (strongly) first order EW phase transition (FOEWPT) to provide the required out-of-equilibrium conditions for baryogenesis in the early Universe~\cite{Sakharov:1967dj}.
Scenarios featuring a FOEWPT have also re-gained attention in recent years since they could lead to a stochastic gravitational wave background detectable with 
future space-based gravitational wave interferometers~\cite{Caprini:2015zlo,Caprini:2019egz}. 

Higgs sector extensions like the two-Higgs-doublet model (2HDM) (see~\cite{Branco:2011iw} for a review) or the Next-to-2HDM (N2HDM)~\cite{Grzadkowski:2009iz,Chen:2013jvg,Drozd:2014yla,Muhlleitner:2016mzt}, which extends the 2HDM by a real scalar singlet field, give rise to a rich collider phenomenology that has an important interplay with
the physics of the early Universe, see e.g.~\cite{Dorsch:2013wja,Dorsch:2014qja,Basler:2016obg,Dorsch:2017nza,Bernon:2017jgv}. 
In this work we explore this interplay within the N2HDM, focusing on type II. We study the thermal history, i.e.\ the evolution of the Higgs fields 
in the early Universe, and demonstrate 
that over large parts of the parameter space of the N2HDM the thermal history in the N2HDM differs very significantly from
the commonly expected scenario of EW symmetry breaking around an early Universe temperature $T$ of $\mathcal{O}(100\gev)$. 
The two phenomena of EW symmetry non-restoration and ``vacuum trapping'' play a key role.

Concerning the feature of EW symmetry non-restoration, it is well-known that
the EW symmetry can be broken already at temperatures much larger than the EW scale, resulting in EW symmetry non-restoration~\cite{Espinosa:2004pn,Meade:2018saz,Baldes:2018nel,Glioti:2018roy,Matsedonskyi:2020mlz, Ramsey-Musolf:2017tgh} up to these (possibly very high) temperatures, or even in no restoration at all. 
In our study of the N2HDM, we find that the presence of this non-restoration behavior 
is related to the effect of the
resummation of infrared divergent modes in the scalar potential within the high-temperature expansion.  We provide compact analytical expressions for the quantities that determine the EW symmetry restoration or non-restoration behavior and supplement our analytical analysis with a detailed numerical investigation.

Vacuum trapping occurs if the Universe remains trapped in an EW symmetric phase down to $T\to 0$, even though a global EW symmetry breaking minimum of the potential exists at zero temperature. We analyze 
this feature in detail for the N2HDM. This effect has also recently been discussed in the context of the NMSSM~\cite{Baum:2020vfl} and also previously in the context of color breaking minima within the MSSM \cite{Cline:1999wi}. Since parameter regions where vacuum trapping occurs are unphysical, we demonstrate that 
the incorporation of this constraint, which up to now has not been taken into account for the
N2HDM, 
has very important consequences for the allowed parameter space of the model.

As a further aspect of
our investigation of the N2HDM thermal history 
we study FOEWPT scenarios in the type~II N2HDM and discuss the interplay between EW symmetry non-restoration, the occurrence of a FOEWPT and vacuum trapping, as well as the connection of such early Universe processes to possible signatures of the N2HDM at the LHC. 
Our results illustrate the rich variety of thermal histories that can be realized in extended Higgs sectors, as well as the phenomenological impact of these different histories. In particular, we demonstrate 
that the results for the thermal history of the early Universe can 
rule out large parts of the otherwise unconstrained N2HDM  
parameter space.

Our paper is organized as follows: In \refse{n2hdmsec} we introduce the N2HDM, paying particular attention to the inclusion of radiative corrections to the scalar potential in \refse{sec:renormalization}, and to the renormalization group evolution of the scalar couplings in section~\ref{sec:rges}. In \refse{constraints} we describe the theoretical and experimental constraints that we take into account for the (zero-temperature) analysis of the N2HDM parameter space. Then, in~\refse{sec:n2hdm-finT} we discuss the inclusion of finite-temperature corrections to the scalar potential and provide a qualitative discussion of the N2HDM thermal history, which we investigate in detail in the following two sections. We study the phenomenon of EW symmetry non-restoration in \refse{section_5_SNR_numerical}, both via an analytical and a numerical approach, and analyze its interplay with the occurrence of a FOEWPT in the N2HDM.
In \refse{section_trapped} we investigate the possible occurrence of vacuum trapping, together with the connection between the thermal history of the N2HDM and its LHC phenomenology.  We conclude
in \refse{conclu}.

\section{The next-to-minimal two Higgs doublet model}
\label{n2hdmsec}
In order to specify our notation and conventions,
we briefly review below the extension of the
CP-conserving (real) 2HDM
with a softly broken $\mathbb{Z}_{2}$ symmetry by a real 
scalar field, the so-called {\it next-to-minimal} 2HDM (N2HDM).
Afterwards, we describe the precise form of
the one-loop zero-temperature effective potential and the renormalization group running of scalar couplings.

\subsection{Model definition and notation}
\label{sec:modeldef}
The tree-level scalar potential of the two $\text{SU(2)}_{L}$ Higgs doublets
$\Phi_{1}$ and $\Phi_{2}$ and the real singlet field $\Phi_{S}$ is given
by~\cite{Muhlleitner:2016mzt}
\begin{align}
V_{\text{tree}}&=m_{11}^{2}\left|\Phi_{1}\right|^{2}+m_{22}^{2}\left|\Phi_{2}\right|^{2}-m_{12}^{2}\left(\Phi_{1}^{\dagger}\Phi_{2}+\text{h.c.}\right)+\frac{\lambda_{1}}{2}\left(\Phi_{1}^{\dagger}\Phi_{1}\right)^{2}+\frac{\lambda_{2}}{2}\left(\Phi_{2}^{\dagger}\Phi_{2}\right)^{2} \notag \\
&+\lambda_{3}\left(\Phi_{1}^{\dagger}\Phi_{1}\right)\left(\Phi_{2}^{\dagger}\Phi_{2}\right)+\lambda_{4}\left(\Phi_{1}^{\dagger}\Phi_{2}\right)\left(\Phi_{2}^{\dagger}\Phi_{1}\right)+\frac{\lambda_{5}}{2}\left[\left(\Phi_{1}^{\dagger}\Phi_{2}\right)^{2}+
\mathrm{h.c.}\right] \notag \\
&+\frac{1}{2}m_{S}^{2}\Phi_{S}^{2}+\frac{\lambda_{6}}{8}\Phi_{S}^{4}+\frac{\lambda_{7}}{2}\left(\Phi_{1}^{\dagger}\Phi_{1}\right)\Phi_{S}^{2}+\frac{\lambda_{8}}{2}\left(\Phi_{2}^{\dagger}\Phi_{2}\right)\Phi_{S}^{2}.\label{Tree-Level Potential}
\end{align}
The $\mathbb{Z}_{2}$ symmetry of the 2HDM potential in~\eqref{Tree-Level
Potential}, $\Phi_{1}\rightarrow\Phi_{1},\,\Phi_{2}\rightarrow-\Phi_{2}$,
whose extension to the Yukawa sector prevents the occurrence of
flavor-changing neutral currents (FCNCs)
at lowest order, is softly broken by the
$m_{12}^{2}$ term.  
The third line of the tree-level potential~\eqref{Tree-Level Potential} includes
the contribution
of the singlet field. Here an extra discrete $\mathbb{Z}'_{2}$ symmetry is
imposed, 
\begin{equation}
\Phi_{1}\rightarrow\Phi_{1} \ , \qquad
\Phi_{2}\rightarrow\Phi_{2} \ , \qquad
\Phi_{S}\rightarrow-\Phi_{S}\label{2ndZ2symmetry} \ ,
\end{equation}
which is not explicitly broken. The original motivation to introduce
this symmetry for the N2HDM was the fact that, when not spontaneously broken,
it will give rise to a dark matter (DM) candidate after
EWSB (see e.g.~\cite{He:2008qm,Grzadkowski:2009iz,Boucenna:2011hy,He:2011gc,Bai:2012nv,Cai:2013zga,Chen:2013jvg,Drozd:2014yla}).
In this work we 
do not restrict to such a scenario but study
the case where $\Phi_{S}$ does
acquire a vacuum expectation value (vev). We 
expand the fields around the EW minimum as follows,
\begin{equation}
\Phi_{1}=\left(\begin{array}{c}
\phi_1^+ \\
\frac{1}{\sqrt{2}}\left(v_{1}+\rho_{1}+i\eta_{1}\right)
\end{array}\right),\quad\Phi_{2}=\left(\begin{array}{c}
\phi_2^+ \\
\frac{1}{\sqrt{2}}\left(v_{2}+\rho_{2}+i\eta_{2}\right)
\end{array}\right),\quad\Phi_{S}=v_{S}+\rho_{3},
\label{Higgs fields expansions}
\end{equation}
where $v_1$, $v_2$ and $v_S$ are the field vevs for the Higgs doublets and the
singlet field, respectively, at zero temperature.
The doublet vevs $v_{1}$ and $v_{2}$ define the EW scale
$v=\sqrt{v_{1}^{2}+v_{2}^{2}} \approx 246 \gev$.
The minimization (or tadpole) equations for $v_1$, $v_2$ and $v_S$ read
\begin{align}
\frac{v_{2}}{v_{1}}m_{12}^{2}-m_{11}^{2}=\frac{1}{2}\left(v_{1}^{2}\lambda_{1}+v_{2}^{2}\lambda_{345}+v_{S}^{2}\lambda_{7}\right)
\ , \label{min_con_1} \\
\frac{v_{1}}{v_{2}}m_{12}^{2}-m_{22}^{2}=\frac{1}{2}\left(v_{1}^{2}\lambda_{345}+v_{2}^{2}\lambda_{2}+v_{S}^{2}\lambda_{8}\right)
\ , \label{min_con_2} \\
-m_{S}^{2}=\frac{1}{2}\left(v_{1}^{2}\lambda_{7}+v_{2}^{2}\lambda_{8}+v_{S}^{2}\lambda_{6}\right),\label{min_con_3}
\end{align}
with $\lambda_{345}\equiv\lambda_{3}+\lambda_{4}+\lambda_{5}.$ 

\vspace{1mm}

Since the CP symmetry and the electric charge are conserved, the (squared-)mass matrix
for the fields $\phi^{\pm}_{1,2}$, $\eta_{1,2}$, $\rho_{1,2,3}$
can be split into three blocks: a $3\times3$ matrix $M_{\rho}^{2}$ for the
CP-even states $\rho_{1,2,3}$,
a $2\times2$ matrix $M_{\eta}^{2}$ for the CP-odd states $\eta_{1,2}$ and
a $2\times2$ matrix $M_{\text{C}}^{2}$ for the charged scalars
$\phi^{\pm}_{1,2}$.
The matrices $M_{\eta}^{2}$ and $M_{\text{C}}^{2}$ correspond to the ones
obtained in the 2HDM, i.e.,
\begin{equation}
M_{\eta,C}^{2}=\frac{m_{A,H^\pm}^{2}}{v^2}
\left(\begin{array}{cc}
v_{2}^{2} & -v_{1}v_{2}\\
-v_{1}v_{2} & v_{1}^{2}
\end{array}\right) \ ,
\end{equation}
with $m_{A}^{2}=v^2 \LP m_{12}^{2}/\LP v_{1}v_{2}\RP - \lambda_{5}\RP$
and $m_{H^\pm}^2 = m_A^2 + v^2 \LP \lambda_5 - \lambda_4 \RP / 2$.
They can be diagonalized via the rotation matrix
\begin{equation}
R_{\beta}=\left(\begin{array}{cc}
c_{\beta} & s_{\beta}\\
-s_{\beta} & c_{\beta}
\end{array}\right) \ ,
\end{equation}
with the abbreviations $s_x \equiv \sin x$ and $c_x \equiv \cos x$,
and 
the angle $\beta$ is defined by 
${t_{\beta} \equiv \tan\beta\equiv v_2/v_1}$.
After diagonalization we are left with the charged and neutral
massless Goldstone bosons $G^{\pm}$ and $G^{0}$
and the charged and neutral CP-odd physical mass
eigenstates $H^{\pm}$ and $A$ with masses $m_{H^\pm}$ and $m_A$.  

The neutral CP-even sector of the N2HDM is modified with respect to that of the
2HDM by the presence of the singlet $\rho_{3}$. The mass matrix $M_{\rho}^{2}$ in
the basis $\rho_{1,2,3}$ can be expressed as
\begin{equation}
M_{\rho}^{2}=\left(\begin{array}{ccc}
v^{2}\lambda_{1}c_{\beta}^{2}+m_{12}^{2}\,t_{\beta} &
v^{2}\lambda_{345}\,c_{\beta}\,s_{\beta}-m_{12}^{2} &
v\,v_{S}\lambda_{7}\,c_{\beta}\\
v^{2}\lambda_{345}\,c_{\beta}\,s_{\beta}-m_{12}^{2} &
v^{2}\lambda_{2}\,s_{\beta}^{2}+m_{12}^{2}/t_{\beta} &
v\,v_{S}\lambda_{8}\,s_{\beta}\\
v\,v_{S}\lambda_{7}\,c_{\beta} & v\,v_{S}\lambda_{8}\,s_{\beta} &
v_{S}^{2}\,\lambda_{6}
\end{array}\right).
\label{eq:massmatrix}
\end{equation}
In the physical basis $h_{1,2,3}$, the mass matrix $M_{\rho}^{2}$ is
diagonal. 
The rotation matrix $R$ between the $h_{1,2,3}$ and $\rho_{1,2,3}$ bases
satisfies 
$R\,M_{\rho}^{2}\,R^{T}=\text{diag}\left(m_{h_{1}}^{2},m_{h_{2}}^{2},m_{h_{3}}^{2}\right)$,
with 
$m_{h_{i}}^{2}$ the squared tree-level mass for $h_i$.
The matrix $R$ can be parametrized in terms of the angles $\alpha_{1,2,3}$ 
\begin{equation}
R=\left(\begin{array}{ccc}
c_{\alpha_{1}}c_{\alpha_{2}} & s_{\alpha_{1}}c_{\alpha_{2}} & s_{\alpha_{2}}\\
-\left(c_{\alpha_{1}}s_{\alpha_{2}}s_{\alpha_{3}}+s_{\alpha_{1}}c_{\alpha_{3}}\right)
&
c_{\alpha_{1}}c_{\alpha_{3}}-s_{\alpha_{1}}s_{\alpha_{2}}s_{\text{\ensuremath{\alpha_{3}}}}
&
c_{\alpha_{2}}s_{\alpha_{3}}\\
-c_{\alpha_{1}}s_{\alpha_{2}}c_{\alpha_{3}}+s_{\alpha_{1}}s_{\alpha_{3}} &
-\left(c_{\alpha_{1}}s_{\alpha_{3}}+s_{\alpha_{1}}s_{\alpha_{2}}c_{\alpha_{3}}\right)
& c_{\alpha_{2}}c_{\alpha_{3}}
\end{array}\right).
\label{mixing matrix}
\end{equation}
Without loss of generality, the angles $\alpha_{1,2,3}$ are defined in the range
$-\pi/2 \leq \alpha_i < \pi/2$, and we choose the convention that  
the mass eigenstates are ordered by ascending mass as
$m_{h_{1}}<m_{h_{2}}<m_{h_{3}}$. 
The singlet composition of the mass eigenstates $h_i$ will be denoted by
$\Sigma_{h_i} = R_{i3}^2$.

The singlet field $\rho_3$ does not couple directly to the SM fermions and gauge
bosons. As a result, any change in the couplings of the CP-even Higgs bosons to
the SM particles w.r.t.\ the ones from the 2HDM is
due to the mixing between the fields $\rho_{1,2}$ and $\rho_3$. 
The Feynman rules for the couplings of the states $h_{i}$ to the massive
gauge bosons $V\equiv W,\,Z$ are 
\begin{equation}
i\,g_{\mu\nu}\,C_{h_{i}VV}\,g_{hVV}^{\text{SM}}\,.
\end{equation}
Here $C_{h_{i}VV}$
are the N2HDM coupling factors of the CP-even Higgs bosons $h_i$ to the massive
SM gauge bosons, and $g_{hVV}^{\text{SM}} $ is the corresponding SM Higgs--gauge
coupling ($g_{hWW}^{\text{SM}} = g \,M_W$, $g_{hZZ}^{\text{SM}} =
\sqrt{g^2+{g'}^2} \,M_Z$).
The coupling factors $C_{h_{i}VV}$ are
given in terms of the mixing matrix elements $R_{ij}$ and the mixing
angle $\beta$ as
\begin{equation}
C_{h_{i}VV} = c_{\beta}R_{i1}+s_{\beta}R_{i2},
\label{eq:effective couplings}
\end{equation}
and, consequently, in terms of the mixing angles $\alpha_{i}$ if we replace
the $R_{ij}$ by their corresponding parametrization shown in \eqref{mixing matrix}.

The $\mathbb{Z}_2$ symmetry in \eqref{Tree-Level Potential} may be extended to
the
Yukawa sector of the theory in order to avoid tree-level 
FCNCs. As the two fields $\Phi_{1}$ and $\Phi_{2}$ transform differently 
under the $\mathbb{Z}_{2}$ symmetry, they cannot be coupled
both to the same SM fermions, which leads to the absence of tree-level
FCNCs. The flavor-conserving Yukawa-types of the N2HDM are those 
of the 2HDM (see e.g.~\cite{Branco:2011iw}).
The Yukawa interactions involving the CP-even Higgs bosons $h_i$ can
be written 
as 
\begin{equation}
\mathcal{L}_{Y}=-\sum_{i=1}^{3}\frac{\sqrt{2} \ m_{f}}{v} \, C_{h_{i}ff} \,
 \bar{\psi}_{f}\,\psi_{f}\,h_{i}\,,
\label{eq: Yukawa lagrangian}
\end{equation}
with the N2HDM coupling factors $C_{h_{i}ff}$ given in Table~\ref{tab:Effective
yukawa couplings}. We also note that any coupling not involving the CP-even neutral Higgs bosons
remains unchanged with respect to the 2HDM and may be found in~\cite{Branco:2011iw}.

\begin{table}
\centering{}
{\renewcommand{\arraystretch}{1.2}
\footnotesize
\begin{tabular}{l||ccc}
 & $u$-type  & $d$-type  & leptons\tabularnewline
\hline 
\hline
Type I  & $R_{i2} / s_{\beta}$  & $R_{i2} / s_{\beta}$  &
    $R_{i2} / s_{\beta}$\tabularnewline
\hline
Type II  & $R_{i2} / s_{\beta}$  & $R_{i1} / c_{\beta}$  &
    $R_{i1} / c_{\beta}$\tabularnewline
\hline
Type III (lepton-specific)  & $R_{i2} / s_{\beta}$ &
    $R_{i2} / s_{\beta}$  & $R_{i1} / c_{\beta}$\tabularnewline
\hline
Type IV (flipped)  & $R_{i2} / s_{\beta}$ & $R_{i1} / c_{\beta}$ &
    $R_{i2} / s_{\beta}$\tabularnewline
\end{tabular}
}
\caption{\small N2HDM coupling factors $C_{h_{i}ff}$ of the CP-even Higgs bosons to
fermions as
defined in \eqref{eq: Yukawa lagrangian}, for the different Yukawa types with no
tree-level FCNCs.}
\label{tab:Effective yukawa couplings}
\end{table}

\subsection{Effective potential and renormalization}
\label{sec:renormalization}

The effective potential (at zero temperature) $V_{\mathrm{eff}}$ incorporates
the effect of radiative corrections into the scalar potential of 
the theory (see e.g.~\cite{Quiros:1999jp} for a review). 
At the one-loop order, $V_{\mathrm{eff}}$ is given by $V_{\mathrm{eff}} = V_{\mathrm{tree}} + V_{\rm
CW}$, where $V_{\rm tree}$ is N2HDM tree-level potential~\eqref{Tree-Level
Potential}, and $V_{\rm CW}$ denotes the well-known Coleman-Weinberg
potential~\cite{Coleman:1973jx}. 
The latter is given in the $\overline{\mathrm{MS}}$
renormalization scheme by
\begin{equation}
V_{\text{CW}}(\phi_i) = \sum_{j}\frac{n_{j}}{64\pi^{2}}(-1)^{2s_{i}} \,
m_{j}(\phi_i)^{4}
\left[\ln\left(\frac{|m_{j}(\phi_i)^{2}|}{\mu^{2}}\right)-c_{j}\right],
\label{CW_potential}
\end{equation}
where $m_{j}(\phi_i)$ is the field-dependent tree-level mass of the particle
species $j$ in our model, $n_{j}$ its corresponding number of degrees of freedom
and $s_{j}$ the particle spin. We here set 
the renormalization scale $\mu$ to be equal to the SM EW vev, $\mu = v$.
The $c_{j}$ are the $\overline{\mathrm{MS}}$ renormalization constants, 
$c_{j}=3/2$ for scalars and fermions and $c_j=5/6$ for gauge bosons. 
For the N2HDM, the sum in~\eqref{CW_potential} runs over
the neutral scalars $\Phi^{0}=\{h_{a},\,h_{b},\,h_{c},\,A,\,G^{0}\}$,
the charged scalars $\Phi^{\pm}=\{H^{\pm},\,G^{\pm}\}$, the SM quarks $q$ and
leptons $\ell$, 
the longitudinal and transversal gauge bosons
$V_{L}=\{Z_{L},\,W_{L}^{+},\,W_{L}^{-}\}$
and $V_{T}=\{Z_{T},\,W_{T}^{+},\,W_{T}^{-}\}$. 
The respective degrees of freedom $n_{j}$ for the species in each category are
\[
n_{\Phi^{0}}=1 \,, \quad n_{\Phi^{\pm}}=2\, ,\quad   n_{V_{T}}=2\, ,\quad
n_{V_{L}}=1\, , \quad n_{q}=12 \,,\quad n_{\ell}=4\, .
\]
The Coleman-Weinberg potential has been evaluated in the Landau
gauge as this allows the omission of ghost contributions from
$V_{\rm CW}$. The effective potential is well-known to be 
gauge dependent, and the extraction of physical information 
from $V_{\mathrm{eff}}$ has to be done with care.\footnote{ 
Often the Nielsen
identities~\cite{Nielsen:1975fs,Aitchison:1983ns} 
are employed in this context; see e.g.~\cite{Patel:2011th} for a
discussion of this issue.}
We note that in the present case the EW symmetry breaking dynamics will be
dominantly dictated by the singlet scalar field, or explored at high $T$
retaining only gauge-invariant contributions. Thus, the gauge dependence of
$V_{\mathrm{eff}}$ is 
of minor concern for our analysis.
For the N2HDM, the sum in \refeq{CW_potential} includes the scalars
$h_{1,2,3}$, $A$, $H^{\pm}$, the Goldstone bosons $G^{\pm}$, $G^{0}$, the
massive EW gauge bosons and the SM fermions (where the main contribution arises from the top quark).

\medskip

The N2HDM tree-level scalar masses and mixing angles differ from those extracted
from the one-loop effective potential if the Coleman-Weinberg potential is
renormalized in the $\overline{\mathrm{MS}}$ scheme. To perform an efficient
scan through the parameter space
of the N2HDM, we have followed~\cite{Basler:2016obg,Basler:2019iuu}
and required that the zero-temperature loop-corrected scalar masses and mixing
angles be
equal to their tree-level values. We achieve this by adding to the effective
potential a UV-finite counterterm contribution  $V_{\rm CT}$, given by 
\begin{equation}
V_{\rm CT}=\sum_{i}\frac{\partial V_{0}}{\partial p_{i}}
\delta p_{i}+\sum_{k}(\phi_{k}+v_{k})\delta T_{k} \ ,
\label{eq:introparacounters}
\end{equation}
where $p_{i}$ stands for the parameters of the tree-level potential. A tadpole
counterterm $\delta T_{k}$ is introduced for each field $\phi_{k}$ which is
allowed to develop a vev. In the case of the
CP-conserving N2HDM the tadpole counterterms are
$\delta T_{1}$, $\delta T_{2}$, $\delta T_{S}$.
They vanish since no additional symmetry is broken
by the radiative corrections at the one-loop
level.
Accordingly, in the following we apply \refeq{eq:introparacounters}
with $\delta T_{k} = 0$.
To maintain the tree-level values of the scalar masses
and their mixing angles at the loop level,
we have imposed the renormalization conditions 
\begin{align}
\partial_{\phi_{i}}V_{\rm CT}(\phi)\left|_{\left\langle
\phi\right\rangle_{T=0}}\right.&=-\partial_{\phi_{i}}V_{\rm CW}(\phi)
\left|_{\left\langle \phi\right\rangle _{T=0}}\right. \ , \\
\partial_{\phi_{i}}\partial_{\phi_{j}}V_{\rm CT}(\phi)
\left|_{\left\langle \phi\right\rangle _{T=0}}\right.&=
-\partial_{\phi_{i}}\partial_{\phi_{j}}V_{\rm CW}(\phi)
\left|_{\left\langle \phi\right\rangle _{T=0}}\right. \ ,
\end{align}
where ${\left\langle \phi\right\rangle _{T=0}}$ corresponds to the tree-level
vacuum state at zero temperature. The derivatives
of the Coleman-Weinberg potential have been computed
following~\cite{Camargo-Molina:2016moz}.
We have found perfect agreement with the implementation of the N2HDM
renormalized zero-temperature effective potential of the public
code~\texttt{BSMPT}~\cite{Basler:2018cwe}.

\subsection{Scale dependence and perturbativity}
\label{sec:rges}
Including radiative corrections leads to a dependence of the
model parameters on the energy scale $\mu$, which is controlled by 
the renormalization
group equations (RGEs).
The running of the parameters can 
drive the couplings into a non-perturbative regime
or even give rise to Landau poles
at scales $\mu > \mu_0$ if the
absolute values of
the quartic couplings $\lambda_i$ at the initial scale~$\mu_0 = v$~are
too large. For the parameter points
that we will discuss in our numerical analysis we verified that
the values $|\lambda_i|$ remain substantially smaller than
the perturbativity bound $4 \, \pi$ over the range
$\mu = [\mu_0, T_{\mathrm{max}}]$, where $T_{\mathrm{max}}$
is the maximum temperature analyzed in each case.

We have numerically solved the RGEs, given in terms of their $\beta$ functions, in the $\overline{\rm MS}$ scheme.
For this analysis we took into account 
the one- and two-loop contributions to the
$\beta$ functions, which we obtained with the help
of the public code
\texttt{SARAH v.4.14.3}~\cite{Staub:2013tta,Schienbein:2018fsw},
solving the general expressions published in
\citeres{Machacek:1983tz,Machacek:1983fi,Machacek:1984zw}.
We have checked the expressions for the $\beta$ functions using
the public code 
\texttt{PyR@TE v.3.0}~\cite{Sartore:2020gou}
and found exact agreement.
For the dominant one-loop terms we analytically checked that
in the limit $\lambda_{6,7,8} \rightarrow 0$
the terms reduce to the ones of the 2HDM, which are well
known in the literature (see e.g.~\cite{Ferreira:2015rha,Basler:2017nzu}).
Due to the renormalization prescription described in section~\ref{sec:renormalization},
which we call ``on-shell'' (OS) in the following,
it is necessary to transform the OS values of the model parameters
$p^{\rm OS}$ at $\mu = \mu_0$ into the corresponding $\overline{\rm MS}$ values
$p^{\overline{\rm MS}}$, such that the
running of the parameters can be applied as described above.
The transformation between the two schemes is given
by the finite parameter counterterms $\delta p_i$
introduced in \refeq{eq:introparacounters}, 
using
\begin{align}
p^{\rm OS}(\mu_0) + \delta p^{\rm OS}(\mu_0) &=
p^{\overline{\rm MS}}(\mu_0)
+ \delta p^{\overline{\rm MS}}(\mu_0) \\
\Rightarrow \ p^{\overline{\rm MS}}(\mu_0) &=
p^{\rm OS}(\mu_0) +
\delta p^{\rm OS}_{\rm fin.}(\mu_0) \ ,
\end{align}
where the second equality follows from the fact that
by definition the counterms $\delta p^{\overline{\rm MS}}$
do not contain finite pieces. 
Accordingly, the counterterms
$\delta p^{\rm OS}_{\rm fin.}(\mu_0)$ 
for the different parameters $p_i$ correspond to
the finite counterterms $\delta p_i$ in \eqref{eq:introparacounters}.

In the perturbative regime, the evolution of the parameters under a variation of $\mu$ is logarithmic.
Therefore, the scale dependence gives rise to only a relatively small uncertainty in the context of
FOEWPTs, which naturally take place at $T \lesssim v $. On the other hand, for the study of the scalar potential at temperatures beyond the EW scale, e.g.\ for the purpose of investigating EW symmetry non-restoration, the variation of the quartic couplings $\lambda_i$ with the energy scale within the whole temperature region can be numerically important. Methods to improve the theoretical uncertainties are discussed e.g.\  in \cite{Croon:2020cgk, Schicho:2021gca}. 
In order to limit the impact of a potentially large 
scale dependence we restrict our analysis to parameter
points with values of $|\lambda_i^{\overline{\rm MS}}(\mu_0)|$
considerably below 
the perturbativity bound $4\pi$ in these cases.
To be more precise, for the benchmark scenarios
discussed in sections~\ref{SNR_framework}
and~\ref{Section_benchmarks} related to non-restoration we
only take into account points with
$|\lambda_i^{\overline{\rm MS}}(\mu_0 = v)| < 3$. 
In section~\ref{SNR_FOEWPT} we discuss the interplay between EW symmetry non-restoration at high $T$ and the occurrence of a FOEWPT at intermediate temperatures. Here, somewhat larger values 
are required in order to give rise 
to a potential barrier between true and false vacuum.
However, the values of $|\lambda_i^{\overline{\rm MS}}(\mu_0)|$
are still substantially smaller than
$4\pi$, and we additionally checked that the
values of 
$|\lambda_i^{\overline{\rm MS}}(\mu)|$
remain below the
bound within the relevant temperature region that has been 
analyzed.
Moreover, the conditions for perturbative unitarity
were applied (see section~\ref{sec:vacstabzerotemp}), yielding further
limitations on 
$|\lambda_i^{\overline{\rm MS}}|$.

\section{Constraints on the N2HDM}
\label{constraints}
As discussed above, the N2HDM has 12 real independent parameters. It is
convenient to choose 
the particle masses of the Higgs sector as input parameters 
since they have a direct physical meaning.
In this section we outline the various theoretical and experimental
constraints that we have imposed and we discuss their impact on the 
parameter space of the N2HDM. For our analysis we have made use of
the public code
\texttt{ScannerS}~\cite{Coimbra:2013qq,Muhlleitner:2020wwk}.
The input parameters that we supply to \texttt{ScannerS} are
\begin{equation}
C^{2}_{h_{a}tt} \ , \; C^{2}_{h_{a}VV}\ , \; \text{sgn}(R_{a3}) \ ,
\; R_{b3} \ ,\; t_{\beta} \ , \; v_{S} \ ,
\; m_{h_{a}} \ , \;m_{h_{b}} \ , \; m_{h_{c}} \ ,
\; m_{A} \ , \; m_{H^{\pm}}\ , \; m_{12}^{2} \ ,
\end{equation}
where the three CP-even scalar mass eigenstates
(not necessarily ordered in mass) are denoted as $h_{a,b,c}$.
We identify these masses with the physical (OS) values; accordingly, the $\lambda_i$ that are obtained from those input values (see~\eqref{Tree-Level Potential}) correspond to on-shell quantities, $\lambda_i^{{\rm OS}}$ (these on-shell parameters are converted to their $\lambda_i^{\overline{{\rm MS}}}$ counterparts where running parameters are required).
Under the assumption $C_{h_{a}VV} \cdot C_{h_{a}tt} >   0$,
the above parameters determine the mixing angles
$\alpha_{1,2,3}$~\cite{Muhlleitner:2020wwk}.
In the following we conveniently choose a parameterization 
where $h_{a}$ is identified with the Higgs boson at about  
$125\,\text{GeV}$, i.e.\
$h_{a}\equiv h_{125}$ and $m_{h_a} \approx 125$ GeV. With an 
appropriate choice of the masses $m_{h_{b,c}}$ the
states $h_{a,b,c}$ can then be identified with the
mass-ordered eigenstates $h_{1,2,3}$.
In the following we briefly describe the constraints
that we have applied in our analysis.

\subsection{Theoretical constraints: Vacuum stability and unitarity}
\label{sec:vacstabzerotemp}
\texttt{ScannerS} discards points in which the tree-level N2HDM scalar
potential is not bounded from below, making use of the conditions
from~\cite{Klimenko:1984qx,Muhlleitner:2016mzt}.
Still, even for a scalar potential that is bounded from below,
the EW minimum might not
be stable (at $T=0$) if it is not
the global minimum of the potential. For such cases,
the probability for the quantum tunnelling from the 
EW minimum into the deeper minima has to be evaluated in order
to determine whether the lifetime of the EW minimum is sufficiently 
long compared to the age of the Universe. This investigation is 
carried out with the linked
public code
\texttt{EVADE}~\cite{Hollik:2018wrr,Ferreira:2019iqb}, which 
determines the minima of the tree-level potential 
and evaluates the lifetime of the EW minimum from the tunnelling 
probability into deeper minima
using a straight path approximation.
Parameter points are regarded as allowed if the 
EW minimum is the global minimum or if it is sufficiently 
long-lived (metastable).

In addition, we require that each parameter point fulfills perturbative
unitarity constraints, formulated in terms of $2 \to 2$ scalar scattering
processes.
We use the approach implemented by default into \texttt{ScannerS} for the N2HDM,
demanding that the eigenvalues of the scattering matrix should 
be smaller than $8 \pi$.
The relevant 
expressions can be found in~\cite{Muhlleitner:2016mzt}. Since various results
discussed in 
sections~\ref{section_5_SNR_numerical} and~\ref{section_trapped} 
involve sizeable quartic scalar couplings, the 
perturbative unitarity
constraints 
play an important role
in our study.

\subsection{Flavor-physics observables}

As discussed in section~\ref{n2hdmsec}, the softly-broken $\mathbb{Z}_2$
symmetry of the 2HDM, 
extended to the Yukawa sector of the theory, forbids FCNCs at tree-level in the N2HDM. 
Accordingly, loop contributions play an important role for the
predictions of low-energy flavor-physics observables, such as rare
$B$-meson decays and $B$-meson mixing parameters, and their
comparison with the experimental results. In most cases,
\footnote{Exceptions are the $B_s \to \mu^+\mu^-$ and
$B_d \to \mu^+\mu^-$ decays, which also depend on the
masses of the additional neutral scalars and the
value of $m_{12}^2$. However, it is known that in the
2HDM type~II these observables are only sensitive to
modifications w.r.t.~the SM for relatively large
values of $\tan\beta \gtrsim 10$~\cite{Haller:2018nnx}.
Since the additional gauge singlet scalar of
the N2HDM does not couple directly to the SM
fermions, the relevant range of $\tan\beta$
is not expected to be substantially modified
compared to the 2HDM. In our analyses we will
use parameter points with $\tan\beta \approx 2$
or below, such that
the constraints from the $B_{s,d} \to \mu^+\mu^-$
decays are of minor importance.}
the dominant deviations from the SM predictions have their
origin in the presence of the charged scalars $H^{\pm}$.
Consequently, in our analysis flavor-physics observables
yield constraints in the $m_{H^\pm}$--$t_{\beta}$ plane, while 
they are relatively insensitive to the remaining parameters of the
N2HDM. This implies that we can safely adopt the flavor
constraints of the 2HDM for our N2HDM analysis.
We have followed the approach implemented in \texttt{ScannerS},
where allowed parameter points are required to be located
within the $2\,\sigma$ region 
of the $m_{H^\pm}$--$t_{\beta}$ plane
as identified via a global fit to experimental data in~\cite{Haller:2018nnx}.
In our analysis the flavor-physics observables exclude values 
of $t_{\beta}\lesssim 0.8$
in all four N2HDM types from Table~\ref{tab:Effective yukawa couplings}.
In addition, a roughly
$t_{\beta}$-independent limit on the charged scalar mass $m_{H^\pm}\gtrsim 600\gev$ is obtained for type~II and IV of the N2HDM.

\subsection{Properties of the observed Higgs boson at 125 GeV}
\label{sechSMconst}

So far, the signal-rate measurements of the Higgs boson
at about 125 GeV that has been discovered at
the LHC agree well with the predictions of the SM. 
For an extended
Higgs sector the compatibility with those measurements requires that
the couplings of $h_a = h_{125}$ should, within the current experimental
uncertainties, resemble
the couplings of a SM Higgs boson. 
In the N2HDM, the effective Higgs couplings (defined as the coupling strength
normalized to the SM prediction for a Higgs boson with the same mass) are
determined by the mixing angles $\alpha_i$ and $t_{\beta}$.
Accordingly, those mixing angles are
constrained by the
LHC Higgs signal-rate measurements.
In order to check the compatibility of the N2HDM parameter space points 
with the
experimentally measured signal rates of the Higgs boson at about 
125 GeV, we use the public
code~\texttt{HiggsSignals
v.2.6.0}~\cite{Bechtle:2013xfa,Stal:2013hwa,Bechtle:2014ewa,
Bechtle:2020uwn},
conveniently linked to \texttt{ScannerS} by default for the N2HDM.
In the scenario of an almost decoupled singlet-like Higgs boson, the N2HDM can
reproduce the alignment and decoupling limits
of the usual 2HDM. In these limits the couplings of $h_a$ 
are equal to the SM couplings.
Since~\texttt{ScannerS} allows one to use $C_{h_a VV}$ and $C_{h_a tt}$ as input
parameters, choosing $C_{h_a VV} \approx 1$,
$C_{h_a tt} \approx 1$ 
yields
parameter points that generally pass the~\texttt{HiggsSignals} test.
A parameter point is regarded to be excluded if
\begin{equation}
\Delta \chi^2_{\rm HiggsSignals} = \chi^2_{\rm N2HDM} - \chi^2_{\rm SM} \geq
6.18 \; ,
\end{equation}
where $\chi^2_{\rm N2HDM}$ and $\chi^2_{\rm SM}$ are calculated via a fit to all
currently available
signal-rate measurements of the Higgs boson at about 125 GeV from 
the Tevatron and the LHC. 
The SM result for a mass of 125 GeV is $\chi^2_{\rm SM} = 84.4$ 
for 107 considered measurements.

\subsection{Direct searches for additional Higgs bosons}

Experimental upper limits on the production of the BSM-type Higgs bosons $h_b, \, h_c$,
$A$ and $H^{\pm}$ provide
important constraints on the parameter space of the N2HDM. 
We take into account the
limits from
Higgs-boson searches at LEP, the Tevatron and the LHC 
for each parameter point by employing the public
code
\texttt{HiggsBounds~v.5.9.0}~\cite{Bechtle:2008jh,Bechtle:2011sb,Bechtle:2013gu,Bechtle:2013wla}.
For the N2HDM, the~\texttt{HiggsBounds} code is linked to~\texttt{ScannerS}
by default, where the so-called effective-coupling input is used for the cross
sections, while the branching ratios are calculated internally using the code
\texttt{N2HDECAY}~\cite{Muhlleitner:2016mzt,Engeln:2018mbg}
and given as input directly to~\texttt{HiggsBounds}. 
The code then 
determines for the considered parameter point the channel with the 
most sensitive expected limit for each Higgs boson and tests whether the
parameter point is allowed at the $95\%$ confidence level by comparing 
for the selected channels the prediction for the production cross section
times branching ratio with the observed upper limit.

\subsection{Electroweak precision observables}

Electroweak precision observables (EWPO) provide constraints on
loop effects arising from the states of the extended Higgs sector 
of the N2HDM. For a BSM model like the N2HDM where only the Higgs sector 
is extended, deviations in the EWPO from the SM can conveniently be
expressed in terms of the
oblique parameters $S$, $T$ and $U$~\cite{Peskin:1990zt,Peskin:1991sw}. 
They are
determined via the gauge-boson self energies 
(see e.g.~\cite{Wells:2005vk} for
more details), and we define them relative to the SM with a Higgs-boson mass of
$\approx 125 \gev$.
The parameter $T$ provides the strongest constraints on the N2HDM parameter
space. Since it accounts for the breaking of custodial symmetry, 
the contributions
of the BSM-type Higgs bosons to $T$ 
approximately vanish 
when either the CP-odd scalar $A$ or the doublet-like
CP-even scalar ($h_{b}$ or $h_{c}$)
are close in mass to the charged scalar $H^\pm$,
i.e.\ $m_A^2 \approx m_{H^\pm}^2$ or $m_{h_{b,c}}^2 \approx m_{H^\pm}^2$.
We make use of the implementation in \texttt{ScannerS} of the $S$, $T$ and $U$
computation for models exclusively containing gauge-singlet and
$SU(2)_L$-doublet scalar fields~\cite{Grimus:2007if,Grimus:2008nb}. 
A point is considered to be excluded if the prediction for the 
$S$, $T$ and $U$ parameters yields a $\Delta\chi^2$ of more than
$2\,\sigma$ relative to the best-fit point from a global fit to the
EWPO~\cite{Haller:2018nnx}.

\section{The N2HDM at finite temperature}
\label{sec:n2hdm-finT}
The discussion of the N2HDM in \refse{n2hdmsec}, followed by the introduction of the relevant
constrains in \refse{constraints}, was limited to the zero-temperature case. In this section, we
will introduce the necessary ingredients to study the N2HDM at finite temperature and analyse the thermal history of the Universe in this scenario.

\subsection{Finite-\texorpdfstring{$T$}{T} effective potential}
\label{sec:sinnlos}
In order to study the thermal history of the N2HDM, one needs to compute the
effective potential including finite temperature corrections. The one-loop
effective potential at finite temperature is given by
\begin{equation}
V\equiv V_{\rm tree} + V_{\rm CW} + V_{T} \, ,
\label{1loopeffectivepotential}
\end{equation}
where $V_{T}$ is the one-loop thermal potential, given
by~\cite{Dolan:1973qd,Quiros:1999jp}
\begin{equation}
V_{T} (\phi_i)=\sum_{j} \, \frac{n_{j} \, T^{4}}{2\pi^{2}}\,
J_{\pm}\left(\frac{m_{j}^2(\phi_i)}{T^{2}}\right)  ,
\label{thermal_potential}
\end{equation}
with the thermal integrals for fermionic ($J_{+}$) and bosonic ($J_{-}$)
particle species 
\begin{equation}
J_{\pm}\left(\frac{m_{j}^2(\phi_i)}{T^{2}}\right) = \mp
\int_{0}^{\infty}dx\,x^{2}\,\log\left[1\pm\exp\left(-\sqrt{x^{2}+\frac{
m_{j}^2(\phi_i)  }{T^{2}}}\right)\right]\, ,
\label{thermalfunctions}
\end{equation}
which vanish as $T\rightarrow 0$ (assuming $m^2_j$ is positive).~In addition to the degrees of freedom 
considered in Eq.~\eqref{CW_potential}, the sum in~\eqref{thermal_potential} includes the photon, since it 
acquires an effective thermal mass at finite temperature (and thus needs to be
included in the sum \eqref{thermal_potential} despite being massless at
${T=0}$).

In certain situations (e.g.~when studying EW symmetry non-restoration at high
$T$, see section~\ref{SNR_framework}) it is
convenient to expand the thermal
functions $J_{\pm}$ in the high temperature 
limit~\cite{Quiros:1999jp}
\begin{eqnarray}
J_{-}(y) &\approx&
-\frac{\pi^{4}}{45}+\frac{\pi^2}{12}y-\frac{\pi}{6}y^{\frac{3}{2}}-\frac{1}{32}\,
y^2 \, \text{log}\left(\frac{|y|}{a_b}\right) + \mathcal{O}(y^3)\, ,  \nonumber \\
J_{+}(y) &\approx&  -\frac{7\pi^{4}}{360}+\frac{\pi^2}{24}y+\frac{1}{32}\, y^2
\, \text{log}\left(\frac{|y|}{a_f}\right) + \mathcal{O}(y^3) \,  \quad \quad
\mathrm{for} \, \left| y \right| \ll 1 \, ,
\label{thermal_expansion}
\end{eqnarray}
with $a_b=\pi^2\text{exp}(3/2-2\gamma_{E})$ and
$a_f=16\pi^2\text{exp}(3/2-2\gamma_{E})$, $\gamma_{E} =  0.57721\ldots$ being the
Euler-Mascheroni constant.
We note that for $m^2_j(\phi) < 0$ both the thermal functions $J_{\pm}$ and the
Coleman-Weinberg 
potential develop imaginary parts. These are related to decay widths of modes
expanded
around unstable regions of field space, as discussed in~\cite{Weinberg:1987vp}
(see also~\cite{Delaunay:2007wb}), and do not have an impact on the 
analysis of the EW phase transition. 
In this work the use of the real part of the effective potential is therefore
implicitly assumed throughout.

A well-known problem of finite-$T$ field theory is the breakdown of the
conventional perturbative expansion and the resulting need to resum a certain
set of higher-loop diagrams~\cite{Gross:1980br,Parwani:1991gq,Arnold:1992rz},
the
so-called daisy contributions (see~\cite{Quiros:1999jp} for a review).
There are several resummation prescriptions in the literature. We here
follow the Arnold-Espinosa method~\cite{Arnold:1992rz}, which amounts to
a resummation of the infrared-divergent contributions from
the bosonic Matsubara zero-modes by adding another piece, $V_{\text{daisy}}$, to
the one-loop effective potential at finite temperature given
in~\refeq{1loopeffectivepotential}. $V_{\text{daisy}}$ is given by
\begin{equation}
V_{\text{daisy}}=-\sum_{k}\frac{T}{12\pi}
\text{Tr}\left[\left(m_{k}^{2}(\phi)+\Pi^2_{k}\right)^{\frac{3}{2}} -  \LP
m_{k}^2(\phi)\RP^{\frac{3}{2}} \right]\  ,
\label{Daisyresummation}
\end{equation}
where the sum in $k$ runs 
over the bosonic degrees of freedom yielding
infrared-divergent contributions, and 
$\Pi^2_k$ denotes their corresponding squared thermal
masses~\cite{Carrington:1991hz}. In the N2HDM, $k$ runs over $W_{L}$, $Z_{L}$,
$\gamma_{L}$ and the field-dependent mass matrices $M_{\text{C}}^{2}(\phi)$,
$M_{\eta}^{2}(\phi)$ and $M_{\rho}^{2}(\phi)$.

Using the Arnold-Espinosa resummation method,
the effective potential
can be treated analytically
in the high-temperature regime
using the expansions of \refeq{thermal_expansion}.
In this limit, the resummation simply amounts to performing
the substitution
$m^2(\phi) \to m^2(\phi) + \Pi^2$ inside the $y^{3/2}$ term 
in~\refeq{thermal_expansion}. 
We have compared 
our resummation prescription with the Parwani resummation
method~\cite{Parwani:1991gq},
also used frequently in the literature. 
Using the Parwani method consists of substituting
$m^2(\phi) \to m^2(\phi) + \Pi^2$
(for the infrared-divergent contributions)
in~\refeq{CW_potential} and~\refeq{thermal_potential}.
The two methods are commonly assumed to be equivalent in the
$m^2(\phi)/T^2 \to 0$
limit, since the field-dependent contributions from the 
logarithmic terms
in~\refeq{CW_potential} and~\refeq{thermal_expansion} cancel each other.
However, it should be noted that the expansion of 
\refeq{thermal_expansion} is no
longer justified when using the Parwani resummation prescription, since $\Pi^2 \sim T^2$ at leading order,
and thus $y = (m^2(\phi) + \Pi^2)/T^2$ does not necessarily go to zero in the
high-$T$ limit. For $m^2(\phi) \ll \Pi^2$, we can use the
expansion
for $J_{\pm} (y)$ from~\cite{Arnold:1992rz,Cline:1996mga,Basler:2016obg},
which includes contributions of $\mathcal{O}(y^3)$
and higher,
to obtain the leading
difference between the two methods in the high-$T$ limit, given by
\begin{eqnarray}
\Delta V_{m^2/T^2 \to 0} &\simeq& \sum_{j\in b} \, \frac{n_{j} \, m_{j}^2(\phi)
T^{2}}{2} \left[ \sum^{\infty}_{\ell = 2} \left(\frac{-\Pi_{j}^2}{4\,\pi^2 T^2}
\right)^{\ell} \frac{(2 \ell -3)!! \, \zeta(2\ell -1)}{(2\ell)!!}\right]
\nonumber \\
&-&  \sum_{j\in f} \, \frac{n_{j} \, m_{j}^2(\phi) T^{2}}{2} \left[
\sum^{\infty}_{\ell = 2} \left(\frac{-\Pi_{j}^2}{4\,\pi^2 T^2} \right)^{\ell}
\frac{(2 \ell -3)!! \, \zeta(2\ell -1)}{(2\ell)!!} (2^{2\ell -1} -1)\right]   \,
,
\label{higher_orders_expansion}
\end{eqnarray}
where $\zeta (x)$ is the Riemman $\zeta$-function, and $(x)!!$ 
denotes the double factorial.
The respective sums are carried out for bosons $b$ and fermions $f$. This
difference can qualitatively modify the high-$T$ behavior of $V$ in specific
regions of parameter space, and even yield a different answer about
the fate of the EW
symmetry in such regions, as we will discuss in more detail in
section~\ref{section_5_SNR_numerical}. 

The leading ($\sim T^2$) contributions to the thermal masses for the scalars in
the N2HDM are given by (in the interaction basis)
\begin{eqnarray}
\label{PI_1}
\Pi_{\rho_{1}\rho_{1}} \hspace{-2mm}
&=&\Pi_{\eta_{1}\eta_{1}}=\Pi_{\phi^+_{1}\phi^+_{1}}=T^{2}\left(c_{1}+\left\{
\begin{array}{rl}
0 \ , & \; \text{Type I/III} \\
\frac{1}{4}y_{b}^{2} \ , & \; \text{Type II/IV}
\end{array}\right.\hspace*{-4pt}\right) , \\
\label{PI_2}
\Pi_{\rho_{2}\rho_{2}} \hspace{-2mm}
&=&\Pi_{\eta_{2}\eta_{2}}=\Pi_{\phi^+_{2}\phi^+_{2}}=T^{2}\left(c_{2}+\left\{
\begin{array}{rl}
\frac{1}{4}y_{b}^{2} \ , & \; \text{Type I/III} \\
0 \ , & \; \text{Type II/IV}
\end{array}\right.\hspace*{-4pt}\right) , \\
\Pi_{\rho_{3}\rho_{3}}\hspace{-2mm}  &=&c_{3}T^{2} ,
\label{PI_3}
\end{eqnarray}
with
\begin{align}
\label{c1} 
c_{1} &=
\frac{1}{16}({g'}^{2}+3g^{2})+\frac{\lambda_{1}}{4}+\frac{\lambda_{3}}{6}+\frac{\lambda_{4}}{12}+\frac{\lambda_{7}}{24}\,
,\\
\label{c2}
c_{2} &=
\frac{1}{16}({g'}^{2}+3g^{2})+\frac{\lambda_{2}}{4}+\frac{\lambda_{3}}{6}+\frac{\lambda_{4}}{12}+\frac{\lambda_{8}}{24}+\frac{1}{4}y_{t}^{2}
\, ,\\
\label{c3}
c_{3} &= \frac{1}{6}(\lambda_{7}+\lambda_{8})+\frac{1}{8}\lambda_{6} \, .
\end{align}
In \refeq{PI_1} and \refeq{PI_2} the only considered fermionic contributions are the
ones from the top and bottom quarks through their respective Yukawa couplings
$y_{t}$ and $y_{b}$. Upon diagonalization of the $M_{k}^2(\phi)+\Pi^2_{k}$
matrices, one can obtain the effective masses including thermal effects for the
N2HDM scalars. The thermal masses of the longitudinal parts
of the SM gauge bosons can be found in~\cite{Basler:2016obg}.

\subsection{N2HDM thermal history}
\label{sec:thermalhistory}

In the following we analyze the thermal history of the N2HDM scalar potential
for the regions of parameter space that satisfy the constraints discussed in
section~\ref{constraints}. 
We use the public code~\texttt{CosmoTransitions}~\cite{Wainwright:2011kj}
to study the scalar potential evolution with temperature, and
analize whether
the Universe evolves to the EW minimum at $T =0$. 
This condition has a highly
non-trivial impact on the physically allowed N2HDM parameter space. 
It implies
that the zero-temperature analysis from section~\ref{constraints} does not
suffice to determine the viable parameter space region of the N2HDM,
since a scalar potential that is
bounded from below and has the EW vacuum as the global minimum at
$T=0$ could still correspond to a scenario that is not physically 
acceptable: 
it is possible that the scalar potential at $T=0$ has more than one
local minimum,
the EW vacuum as global minimum and, for instance,
a metastable vacuum with 
$\left\langle \Phi_{1,2}
\right\rangle = 0$, $\left\langle \Phi_{S} \right\rangle \neq 0$. 
If at some
temperature $T > 0$ only the $\left\langle \Phi_{1,2} \right\rangle = 0$,
$\left\langle \Phi_{S} \right\rangle \neq 0$ vacuum is present,
the Universe can
only evolve to the EW minimum by tunnelling from the metastable one.
Then, if the
corresponding tunnelling probability is never large enough to permit the
transition, the Universe would be trapped in the
metastable vacuum at ${T=0}$.

Previous studies of the N2HDM in the early
Universe~\cite{Basler:2018cwe,Basler:2019iuu} have relied on identifying the
critical temperature $T_c$ at which the EW minimum would have been degenerate in
energy with other potential vacua.\footnote{We would like to
stress that the definition of the critical
temperature here (in the context
of first-order phase transitions) should not be confused with the
definition of the critical temperature
in the context of second-order (continuous)
phase transitions, in which it usually refers to
the temperature at which the transition occurs.}
Therein, it was furthermore assumed that the phase
transition to the EW vacuum always takes place if the EW vacuum is the global
minimum of the potential at $T =0$. However, as argued above, this is by no
means guaranteed, but depends on the false vacuum tunnelling rate per unit time
and volume~\cite{Coleman:1977py,Callan:1977pt,Linde:1980tt,Linde:1981zj} 
\begin{equation}
\Gamma(T)=A(T)\,{\rm e}^{-S_{3}(T)/T},
\label{eqfuncdeter}
\end{equation}
where $S_3$ denotes 
the three-dimensional action for the ``bounce'' (multi-)field
configuration $\phi_{\rm{B}}$ that interpolates between the metastable 
vacuum and the EW vacuum 
for $T < T_c$, 
\begin{equation}
S_{3} = 4\pi \int r^{2} {\rm d}r \, \left[\frac{1}{2}
\left(\frac{{\rm d}\phi_{\rm{B}}}{{\rm d}r}\right)^{2}+
V \left(\phi_{\rm{B}},T\right)\right]\, .
\label{eq:bounceaction}
\end{equation}
The bounce $\phi_{\rm{B}}$ is the 
configuration of scalar fields that extremizes the
action given in \refeq{eq:bounceaction} with the boundary conditions
that $\left. {\rm d}\phi/{\rm d}r\right|_{r=0}=0$ holds and 
the false vacuum is approached for
$r \to \infty$. 
The prefactor $A(T)$ is a functional determinant~\cite{Callan:1977pt} given
approximately by $A(T) \approx T^4 \, (S_3/2\pi T)^{3/2}$~\cite{Linde:1980tt}.
The onset of the phase transition requires that the time integral of the
tunnelling rate of \refeq{eqfuncdeter} in a Hubble volume $H$, which can be
correspondingly expressed as a temperature integral, is $\approx 1$
(see e.g.~\cite{Espinosa:2008kw}). This defines the nucleation temperature
$T_{n}$,
\begin{equation}
\int_{T_n}^{T_c} \frac{T^4}{H^4} \frac{A(T)}{T} \,{\rm e}^{-S_{3}(T)/T}\, {\rm d}T \approx 1 
\, .
\label{eq:_TN}
\end{equation}
Here the Universe is assumed to be dominated by radiation, and the Hubble
parameter $H$ is given by
$H^2 = (8\,\pi^3 g_{\star} \, T^4)/(90\, \rm{M}_{\rm{Pl}}^2)$, where
$g_{\star}$ denotes
the effective number of relativistic degrees of freedom, and 
$\rm{M}_{\rm{Pl}}=
1.22 \times 10^{19} $ GeV is the Planck mass.
If the condition of \refeq{eq:_TN} is not satisfied for any 
temperature below
$T_c$, then the Universe is trapped in the metastable vacuum, and EW symmetry
breaking never 
occurs
(see e.g.~\cite{Espinosa:2008kw} for a discussion).\footnote{Even
if such a metastable vacuum
eventually decays through quantum
tunnelling~\cite{Coleman:1977py}, such a situation yields an inflationary scenario
with a ``{\it graceful-exit}'' problem~\cite{Guth:1982pn}, 
where successful reheating after inflation is not achieved 
and therefore the observed
Universe cannot be reproduced}
As will be discussed below, we find 
that such a situation is quite common in the N2HDM,
with the metastable vacuum corresponding to
a minimum in which only $\left\langle \Phi_{S} \right\rangle$
is non-zero.
\footnote{
A corresponding observation of such a situation 
has recently been made for the
NMSSM~\cite{Baum:2020vfl}.}
In particular, when aiming to identify the regions of the N2HDM parameter space
where a 
first order EW phase transition is possible, the approach based 
just on
$T_c$~\cite{Basler:2018cwe,Basler:2019iuu} (but not $T_n$) 
is not sufficient and can result in rather misleading specifications of
the parameter space where such a 
first order phase transition is realized.
In section~\ref{section_trapped} we will 
discuss these ``trapped-vacuum'' scenarios in detail. 

In addition, our study of the thermal history of the N2HDM reveals the
possibility that the EW symmetry is not restored at high $T$, as well as the
possible
non-restoration of the discrete $\mathbb{Z}'_2$ symmetry
of the singlet sector,
see \refeq{2ndZ2symmetry}. The possibility of  $\mathbb{Z}'_2$  symmetry non-restoration at high temperature has also been explored for a singlet extension of the SM  \cite{Carena:2019une}.
Scenarios where the EW symmetry is not restored around
the EW scale, and
where the
restoration is either pushed above the TeV scale or
 does not occur at all, have
recently regained attention~\cite{Meade:2018saz,Baldes:2018nel,Glioti:2018roy,
Matsedonskyi:2020mlz}.
They lead to important consequences for the generation of
the matter-antimatter asymmetry in the early Universe: the EW sphaleron
processes, which constitute the source of baryon (and lepton) number violation
in the SM, would not be active at $\mathcal{O}(100\, \rm{GeV})$ temperatures,
and could have even been suppressed for the entire thermal history of the
Universe. We will discuss the details of EW symmetry non-restoration within the N2HDM
in the next section.

\section{Symmetry non-restoration at high \texorpdfstring{$T$}{T}}
\label{section_5_SNR_numerical}
We now investigate whether the EW symmetry remains un-restored at high $T$
within the N2HDM. In a first step we do this analytically 
by studying the curvature of the
effective potential around $\left\langle \Phi_{1} \right\rangle = 0$,
$\left\langle \Phi_{2} \right\rangle = 0$ in the high-temperature limit.
We find that,
under certain assumptions (see below),
the fate of the EW symmetry at high temperatures (restoration \textsl{vs.} non-restoration) can
be reliably determined from our analysis, while this is not necessarily true for the
restoration or non-restoration of the $\mathbb{Z}_{2}'$ symmetry of 
the singlet field.  We then compare our analytical results with our
numerical study of the effective potential evolution with temperature in section~\ref{Section_benchmarks}, and discuss the implications of our results for the EW phase transition in section~\ref{SNR_FOEWPT}.

\subsection{Analytical considerations}
\label{SNR_framework}

In order to analytically study the behavior of the effective potential at high $T$,  
we use~\refeq{thermal_expansion} for the thermal functions $J_{\pm}$ and compute
$V_{\mathrm{daisy}}$, given by~\refeq{Daisyresummation}, in the limit
$m^2(\phi)/T^2 \ll 1$.
In addition, since the leading ($\sim T^2$) contributions to the squared thermal
masses $\Pi^2$ enter only into the diagonal elements of the scalar mass
matrices, as shown in \refeqs{PI_1}--\eqref{PI_3},
the off-diagonal terms can be neglected in the high-$T$ limit.

The restoration of \textit{both} the EW symmetry and the discrete $\mathbb{Z}_{2}'$ 
symmetry of the singlet field 
requires the origin of field space $\left\langle \Phi_{1}
\right\rangle = 0$, $\left\langle \Phi_{2} \right\rangle = 0$, $\left\langle
\Phi_{S} \right\rangle = 0$ to be 
a minimum at high temperature. 
In order to assess whether this is the case we
compute the principal minors of the Hessian matrix 
$H^0_{ij} = \left. \partial^2 V / \partial \rho_i \partial \rho_j
\right|_{(0,0,0)}$
as a function of the parameters of the theory.
The conditions for the origin to be a  minimum of the N2HDM potential at
large $T$ (large in comparison to the bilinear terms of the theory) are
\begin{align}
H_{11}^{0}&>0 \ ,
\label{v1_derivative} \\
H_{11}^{0}H_{22}^{0}-\left(H_{12}^{0}\right)^2&>0 \ ,
\label{minor_determinant} \\H_{33}^{0}&>0 \ , 
\label{hessian_determinant}
\end{align}
where we made use of the fact that $H_{13}^0 = H_{23}^0 = 0$.
Since the cross derivative $H_{12}^{0} = -m_{12}^2$ does not depend on $T$, the 
above conditions can be simply cast as $c_{ii} \equiv
\lim\limits_{\,T\rightarrow\infty}\,
H_{ii}^{0}/ T^2 > 0$ with $i = 1,2,3$. The coefficients
$c_{ii}$ are given by
\begin{align}
c_{11} & \simeq
-0.025+c_{1}-\frac{1}{2\pi}\left(\frac{3}{2}\lambda_{1}\sqrt{c_1}+\lambda_{3}\sqrt{c_{2}}+\frac{1}{2}\lambda_{4}\sqrt{c_{2}}+\frac{1}{4}\lambda_{7}\sqrt{c_{3}}\right)
\ ,
\label{coeff_1}
\\
c_{22} & \simeq
-0.025+c_{2}-\frac{1}{2\pi}\left(\frac{3}{2}\lambda_{2}\sqrt{c_2}+\lambda_{3}\sqrt{c_{1}}+\frac{1}{2}\lambda_{4}\sqrt{c_{1}}+\frac{1}{4}\lambda_{8}\sqrt{c_{3}}\right)
\ ,
\label{coeff_2}
\\
c_{33} & =
c_{3}-\frac{1}{2\pi}\left(\lambda_{7}\sqrt{c_1}+\lambda_{8}\sqrt{c_{2}}+\frac{3}{4}\lambda_{6}\sqrt{c_{3}}\right)\, ,
\label{coeff_3}
\end{align}
with $\lambda_i \equiv \lambda_i^{{\rm OS}}$.
Here, the contribution of the SM gauge couplings $g$ and $g'$ to~\eqref{coeff_1}
and~\eqref{coeff_2} arising from the resummation of daisy diagrams is given
numerically ($\approx - 0.025$) for reasons of compactness.
Even though our analysis focuses in the N2HDM type-II, these coefficients are valid for all Yukawa types up to subleading corrections proportional to the tau lepton and the bottom quark Yukawa couplings.
The quantity $c_{2}$ receives a large positive
contribution from the top Yukawa coupling
(see~\refeq{c2}).
Thus, for the moderate values of $\lambda_{i}$ 
used in our analysis one finds $c_{22}>c_{11}$,
and accordingly the
simultaneous restoration of both the EW and  $\mathbb{Z}_{2}'$ symmetries
at high temperature occurs for positive $c_{11}$ and $c_{33}$. 
In contrast, for positive $c_{33}$ but negative $c_{11}$ the EW symmetry is not restored at high temperatures.\footnote{We note that in this case, determining whether the $\mathbb{Z}_{2}'$ symmetry is restored or non-restored at high $T$ would require the exploration of the N2HDM scalar potential away from the origin of field space, along the EW field directions (for $\left\langle \Phi_{1}
\right\rangle \neq 0$, $\left\langle \Phi_{2} \right\rangle \neq 0$), 
corresponding to a much more involved analysis.}

For $c_{33}<0$, the origin of field space is unstable along the singlet field direction $\rho_3$. The analysis of EW symmetry restoration in this case requires the investigation of the curvature of the effective potential at high temperature around $\left\langle \Phi_{1} \right\rangle = 0$, $\left\langle \Phi_{2} \right\rangle = 0$, $\left\langle
\Phi_{S} \right\rangle = v_S(T)$, where $v_S(T)$ denotes the (nonzero) minimum of the potential along the $\rho_3$ field direction. 
The curvature around $(0,\,0,\,v_S(T))$ in the direction of $\rho_1$ is given by
\begin{equation}
    c_{11}^S=\lim\limits_{T\rightarrow\infty} \frac{H_{11}^{S}(v_S(T),T)}{T^2}\,\, , \mbox{~~~with~~~} 
    H_{11}^S(\rho_3,T) =\left. \frac{\partial^2V}{\partial\rho_1^2}\right|_{(0,0,\rho_3)} \ .
\end{equation}
Then, $c_{11}^{S}<0$ is a sufficient condition for EW symmetry non-restoration when $c_{33}<0$. As discussed above, from the large positive contribution of the top Yukawa coupling to the curvature in the direction of $\rho_2$, we generally expect this to be larger than the curvature in the direction of $\rho_1$. So, for $c_{11}^{S}>0$ (when $c_{33}<0$) the EW symmetry is generally restored at high $T$. The coefficient $c_{11}^{S}$ takes the form
\begin{align}
c_{11}^S &=
\lim\limits_{T\rightarrow\infty} \Bigg\{
-0.025+c_{1}+\frac{\lambda_{7}}{2}\frac{v_{S}^{2}(T)}{T^{2}}
-\frac{1}{2\pi}\Bigg(\frac{3}{2}\lambda_{1}\sqrt{c_{1}+
\frac{\lambda_{7}}{2}\frac{v_{S}^{2}(T)}{T^{2}}}
\notag \\
&+\lambda_{3}\sqrt{c_{2}+
\frac{\lambda_{8}}{2}\frac{v_{S}^{2}(T)}{T^{2}}}+
\frac{\lambda_{4}}{2}\sqrt{c_{2}+
\frac{\lambda_{8}}{2}\frac{v_{S}^{2}(T)}{T^{2}}}
+\frac{\lambda_{7}}{4}\sqrt{c_{3}+
\frac{\lambda_{3}}{2}\lambda_{6}
\frac{v_{S}^{2}(T)}{T^{2}}} \Bigg) \Bigg\}
\notag \\
&= c_{11} +
\mathcal{O}\left(\frac{v_S(T)^{2}}{T^{2}}\right)\, .
\label{H11S_limit}
\end{align}
From a computational perspective, calculating $c_{11}^S$ is slightly more involved than obtaining $c_{11}$ since one has to identify the extrema of the scalar potential in the plane $(0,0,\rho_{3})$ as a function of temperature to obtain $v_{S}(T)$.
We also remark that the analysis of EW symmetry non-restoration based on the sign of $c_{11}^S$ relies on the validity of the 
high-$T$ expansion: the
N2HDM scalar masses evaluated at $(0,\,0,\,v_{S}(T))$ receive contributions (dependent on $\lambda_{6}$, $\lambda_{7}$ and $\lambda_{8}$) proportional to the singlet vev $v_S(T)$, and $|v_S(T)|$ will be a monotonically increasing function of temperature. In order to guarantee that these contributions 
do not render the scalar masses at $(0,\,0,\,v_{S}(T))$ comparable to the temperature, thus invalidating the high-$T$ expansion, we require $|\lambda_{6}|,
|\lambda_{7}|, |\lambda_{8}| < 1$
at the initial scale $\mu_0 = v$.\footnote{The
dependence of $\lambda_{6,7,8}$ on $\mu$ is very mild
for $|\lambda_{6,7,8}(\mu_0)| < 1$ due to the
singlet nature of $\rho_3$.
In the following, all quoted values
of $\lambda_{6,7,8}$ (e.g.~in Table~\ref{tab:coefficients_examples_analytic}) are understood
to be given at $\mu_0 = v$, having in mind that
they are not substantially different
at $\mu > v$ within the perturbative regime.}

From \refeq{H11S_limit}, if the corrections proportional to $v_{S}(T)^{2}/T^{2}$ are subleading compared to 
the coefficient $c_{11}$, i.e.~$|c_{11}| \gg |v_S(T)^2/T^2|$, this coefficient $c_{11}$ 
defined at the origin in field space also controls the stability of the field space point $(0,0,v_{S}(T))$ 
in the direction of $\rho_{1}$ in the high-temperature limit. Then, the sign of $c_{11}$ 
determines the high-$T$ restoration/non-restoration of the EW symmetry for both 
$c_{33} > 0$ and $c_{33} < 0$.
On the other hand, if the $\mathcal{O}\left(v_{S}(T)^{2}/T^{2}\right)$ term in \refeq{H11S_limit} is comparable in size to $c_{11}$, then the full calculation of $c^S_{11}$ is needed to assess the fate of the EW symmetry at high $T$ (when $c_{33} < 0$). The coupling $\lambda_{7}$ plays an important role in this respect:
the only $\mathcal{O}(v_S(T)^2 / T^2)$ term in Eq.~\eqref{H11S_limit} proportional to a single power of $\lambda_i$, and not suppressed by an additional $(2 \pi)^{-1}$ factor, depends precisely on $\lambda_7$. 
This is then the most important parameter for the $\mathcal{O}(v_S(T)^2 / T^2)$ corrections in $c_{11}^S$. 

\begin{table}
\centering
{\renewcommand{\arraystretch}{1.4}
\footnotesize
\begin{tabular}{c||cccccccccccc}
& $m_{h_{1}}$  & $m_{h_{2}}$  & $m_{h_{3}}$ & $m_{A}$ & $m_{H^{\pm}}$  &
$t_{\beta}$ & $C_{h_{1}tt}$ &  $C_{h_{1}VV}$ &
$\text{sgn}\left(R_{13}\right)$  & $R_{23}$  & $m_{12}^2$ & $v_{S}$ \\
\hline
\hline
A$_1$ & $125.09$  & $934$  & $1263$ & $1008$ & $958$ & $1.72$ & $0.94$ & $0.94$ &
$-1$ & $-0.22$ & $604^2$ & $2637$\\
\hline 
A$_2$ & $125.09$  & $840$  & $1355$ & $904$ & $828$ & $1.73$ & $0.99$ & $0.96$ &
$-1$ & $-0.104$  & $557^2$ & $2298$ \\
 \hline 
B$_1$ & $125.09$  & $589$  & $760$ & $739$ & $748$ & $1.51$ & $0.99$ & $0.99$ &
$+1$ & $-0.96$ & $495^2$ & $2500$\\
\hline
B$_2$ & $125.09$  & $685$  & $700$ & $680$ & $678$ & $2$ & $0.97$ & $0.97$ & $+1$ &
$-0.46$  & $436^2$ & $1100$ \\
 \hline 
C$_1$ & $125.09$  & $835$  & $1370$ & $897$ & $834$ & $1.1$ & $0.97$ & $0.96$ &
$+1$ & $0.04$ & $559^2$ & $2707$\\
\hline
C$_2$ & $125.09$  & $792$  & $850$ & $835$ & $814$ & $1.02$ & $0.99$ & $0.99$ & $-1$
& $0.51$  & $510^2$ & $2565$ \\
\hline 
D & $125.09$  & $408$  & $717$ & $731$ & $707$ & $2$ & $0.99$ & $0.99$ &  $+1$ &
$0.86$ & $380^2$ & $1487$\\

\end{tabular}
}
\caption{\small Illustrative type-II N2HDM benchmarks for high-$T$ EW 
symmetry restoration/non-restoration, in terms of \texttt{ScannerS} input parameters. The
parameters $m_{h_i}$, $m_A$, $ m_{H^{\pm}}$, $m_{12}$ and $v_S$ are given in
GeV.}
\label{tab:input_parameters_benchmark_physical_2}
\end{table}

In order to illustrate this analytic assessment of the EW symmetry non-restoration behavior at high temperature, we now discuss several benchmark scenarios (A$_{1,2}$, B$_{1,2}$, C$_{1,2}$, D), defined in Table~\ref{tab:input_parameters_benchmark_physical_2} in terms of their \texttt{ScannerS} input parameters, which are in agreement with all constraints discussed in section~\ref{constraints}.
We have required $|\lambda_{6,7,8}| < 1$ 
for all benchmarks to ensure the validity of the high-$T$ expansion, and imposed $|\lambda_{1,\dots,5}(\mu = v)| < 3$ for the other quartic couplings to guarantee that they remain perturbative much above the TeV scale, as discussed in section~\ref{sec:rges}. The bounds on $|\lambda_i|$ lead to a common feature for all benchmarks: the pseudoscalar $A$, the charged Higgs bosons $H^\pm$ and the heavy doublet-like scalar
are close to each other in mass, with their mass scale
roughly given by $M \equiv\sqrt{m_{12}^2 /
(s_\beta c_\beta)}$.\footnote{This is $h_2$ for
benchmarks A$_{1,2}$, C$_{1}$ and $h_3$
for benchmarks B$_{1}$, D.
For benchmarks B$_{2}$, C$_{2}$ the
doublet--singlet mixing is 
sizeable, such that $m_{h_{2,3}} \approx M$.}
Also, the values $t_\beta \gtrsim 1$
that have been chosen for all displayed benchmarks correspond to the parameter region for which the various theoretical and experimental constraints are most easily accommodated~\cite{Muhlleitner:2016mzt}.
In Table~\ref{tab:coefficients_examples_analytic} we show the values of $c_{11}$ and $c_{33}$ for each of the benchmarks (we have verified that $c_{22} > 0$ for all of them). For benchmarks C$_{1,2}$ we find $c_{33} > 0$, and accordingly the sign of $c_{11}$ fully determines the fate of EW symmetry at high temperature. In both cases $c_{11} < 0$ holds, and thus the EW symmetry is un-restored at high $T$. For benchmarks A$_{1,2}$, B$_{1,2}$ and D we have $c_{33} < 0$, and thus the origin of field space is unstable along the singlet field direction. The possible restoration of the EW symmetry in this case is controlled by the sign of $c^S_{11}$, also shown in Table~\ref{tab:coefficients_examples_analytic}. For B$_{1,2}$ and D we find $c^S_{11} > 0$, and thus the EW symmetry is restored at high $T$, while the singlet $\mathbb{Z}_{2}'$ symmetry remains broken at high $T$. In contrast, for A$_{1,2}$ the EW symmetry is not restored at high temperature since $c^S_{11} < 0$.

The scenarios A$_{1,2}$ and B$_{1,2}$ are benchmarks for which $|c_{11}| \gg |v_S(T)^2/T^2|$ at high temperature, such that $c_{11}$ determines the fate of the EW symmetry in this limit. The signs of $c_{11}$ and $c_{11}^{S}$ are the same for such a case,
as shown explicitly in 
Table~\ref{tab:coefficients_examples_analytic}. 
In contrast, for benchmark D we have $c_{11}^S > 0$ despite the negative value of $c_{11}$. 
This behavior is caused by the small value of $|c_{11}|$ (the smallest among all benchmarks)
together with a sizeable value of $\lambda_{7}$ (the largest among all benchmarks, also shown in
Table~\ref{tab:coefficients_examples_analytic}).
This renders the contribution given by $\lambda_7/2\; v_{S}^2(T)/T^{2}$ in~\eqref{H11S_limit} large in comparison to $c_{11}$, leading to the restoration of the EW symmetry at high $T$ even for $c_{11} < 0$.

\begin{table}
\centering
{\renewcommand{\arraystretch}{1.4}
\footnotesize
\begin{tabular}{l||ccccccc}
 & A$_1$ & A$_2$ & B$_1$ & B$_2$ & C$_1$ & C$_2$ & D \\
\hline
\hline
$c_{11}$ & -0.092 & $-0.06$ & 0.182 & $0.10$ & -0.104 & $-0.04$  & -0.006 \\
\hline
$c_{33}$ & -0.011 & $-0.02$ & -0.002 & $-0.002$ & 0.058 & $0.005$  & -0.010 \\
\hline
$\mathrm{sgn}(c_{11}^{S})$ & - & - & + &  + & - & -  & + \\
\hline
\hline
$\lambda_6$ & 0.211 & 0.329    & 0.058 &  0.382   & 0.246 &  0.104    & 0.115 \\
\hline
$\lambda_7$ & 0.154 & 0.400   &  -0.199  &   -0.440   & -0.465 &  0.218   & 0.760 \\
\hline
$\lambda_8$ & 0.703 &  0.986    & 0.007 &   -0.362    & -0.613 &   0.087    & 0.271
\end{tabular}
}
\caption{\small The values of $c_{11}$,
$c_{33}$ and $\mathrm{sgn}(c_{11}^{S})$ for
each of the benchmarks defined in 
Table~\ref{tab:input_parameters_benchmark_physical_2}.  
For all displayed benchmark scenarios $c_{22} > 0$ holds.
Also shown are the values of the singlet field
quartic couplings $\lambda_{6,7,8} =
\lambda_{6,7,8}^{\rm OS}$.}
\label{tab:coefficients_examples_analytic}
\end{table}

\begin{figure}[t!]
\centering
\includegraphics[width=0.42\textwidth]{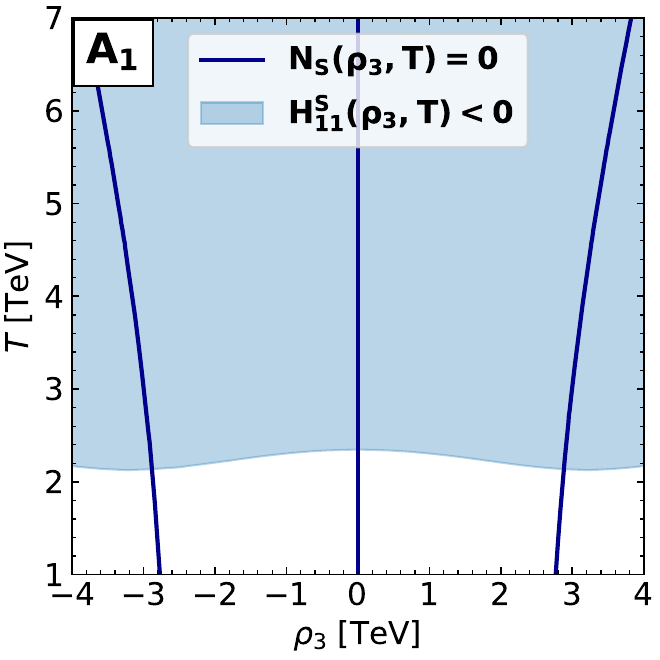}
\hspace{0.01\textwidth}
\includegraphics[width=0.42\textwidth]{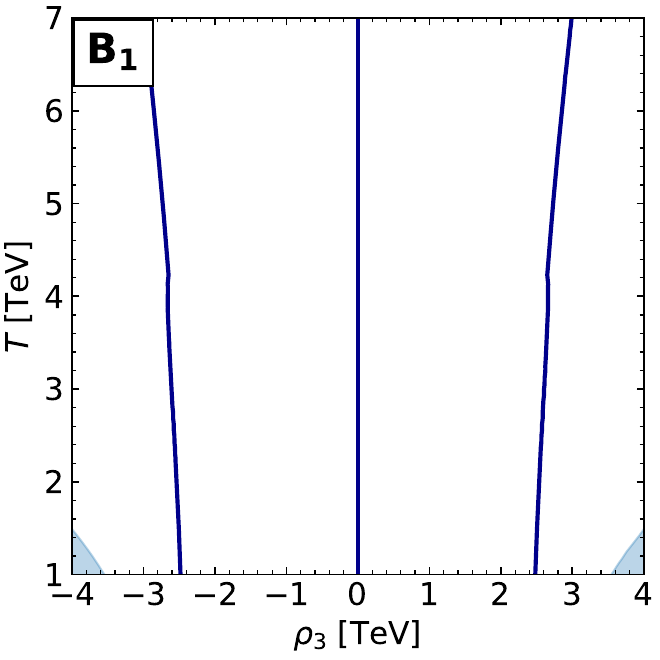}\\
\vspace{0.002\textwidth}
\includegraphics[width=0.42\textwidth]{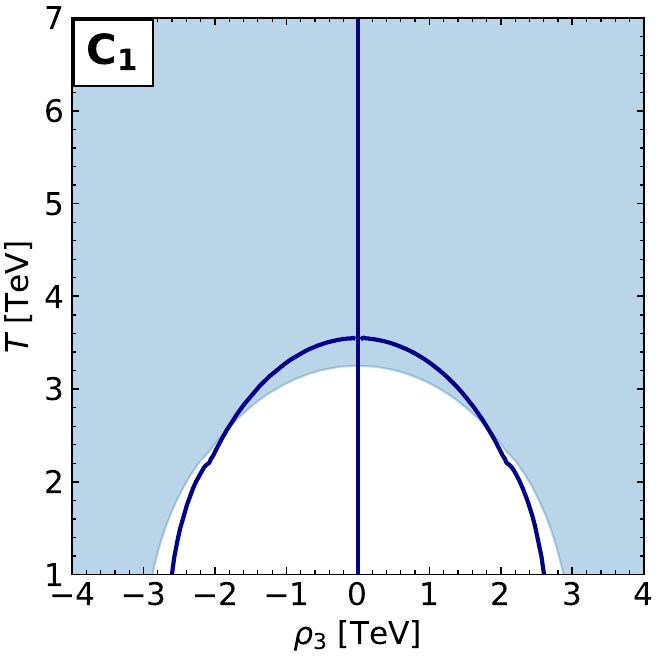}
\hspace{0.01\textwidth}
\includegraphics[width=0.42\textwidth]{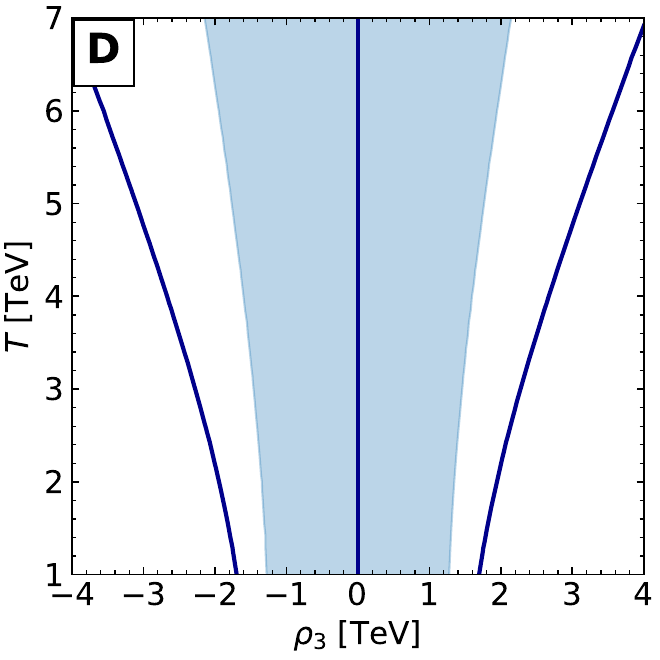}
\caption{\small ($\rho_{3}$--$T$) plane for four qualitatively different benchmark scenarios (depending on the sign of the coefficients $c_{11}$, $c_{33}$ and $c_{11}^S$, see text for details) from Table~\ref{tab:input_parameters_benchmark_physical_2}, A$_1$ (top-left), B$_1$ (top-right), C$_1$ (bottom-left) and D (bottom-right). The region for which $H_{11}^{S}(\rho_{3}, T)<0$ is depicted in light blue. The dark blue lines indicate the stationary points with $N_{S}(\rho_{3},T) = 0$ 
of the high temperature approximation of the potential. 
}
\label{c_ii_analytical}
\end{figure}

The different types of scenarios regarding EW symmetry restoration or non-restoration at high temperature discussed above are illustrated in Figure~\ref{c_ii_analytical}, where each plot corresponds to a different type of benchmark (A, B, C, D). Figure~\ref{c_ii_analytical} shows the behavior of the effective potential (in the high $T$ approximation) along the $(0,\, 0,\, \rho_3)$ field space direction and in dependence of $T$. The region for which $H_{11}^{S}(\rho_{3},T)<0$ is depicted in light blue, and 
the dark blue lines show the stationary points along the singlet field
direction,\footnote{Note that along the
$\rho_1$ and $\rho_2$ field directions
the derivatives are $0$ automatically
for ($0,0,\rho_3$) due to $SU(2)_L$ gauge
invariance.} i.e.~the solutions to
\begin{equation}
N_{S}(\rho_{3},T)=\left.\frac{\partial V(T)}{\partial \rho_3}\right|_{(0,0,\rho_3)}=0\, .
\end{equation}
Given the symmetry of the potential, these solutions correspond to $\rho_3 = 0$ and $ \rho_3 = \pm v_S (T)$, the latter only appearing as solutions (in this case, yielding two identical stationary points) when the field space point ($0,\, 0,\, 0$) is either a maximum or a saddle point of the effective potential. When a dark blue line in Figure~\ref{c_ii_analytical} lies within the light blue region, the corresponding extremum along the singlet direction $\rho_3$ is unstable along the $\rho_1$ field direction, and the EW symmetry will not be restored there.
For benchmark A$_1$ in the upper-left plot of Figure~\ref{c_ii_analytical}, none of the $N_{S}(\rho_{3},T)$
stationary points is stable in the direction of $\rho_{1}$ for $T \gtrsim 2$ TeV. Therefore, the EW symmetry is inevitably un-restored at high temperature.
In contrast, for benchmark B$_1$ in the upper-right plot of Figure~\ref{c_ii_analytical}, the extrema 
${(0,0,\pm v_S(T))}$ are stable along the $\rho_{1}$ field direction and correspond to global minima of the N2HDM potential at high temperature, leading to EW symmetry restoration. 
For benchmark C$_1$ in the lower-left plot of Figure~\ref{c_ii_analytical}, the only extremum along the singlet direction at high $T$ (in this case, for $T \gtrsim 3.5\tev$) is $\rho_3 = 0$, since $ c_{33} > 0$. Yet, the origin of field space is unstable in the direction of $\rho_1$ at high $T$, as a result of
$H_{11}^{S}(0,T \gtrsim 3\,\mathrm{TeV}) < 0$, and the EW symmetry is therefore not restored in this case.
Finally, for benchmark D in the lower-right plot of Figure~\ref{c_ii_analytical}, we observe that for the extremum $\rho_3 = 0$ (a maximum along the singlet direction), we have $H_{11}^{S}(0,T) < 0$ as a consequence of $c_{11} < 0$. However, for the other two extrema $\rho_3 = \pm v_S (T)$, which correspond to minima along the singlet direction, we have $H_{11}^{S}(\pm v_S (T),T) > 0$. This leads to EW symmetry restoration at high temperature in the field space points ($0,0,\pm v_S(T)$).

\vspace{2mm}

To summarize, our analytical approach based on the high-$T$ expansion of the effective potential allows one to determine the restoration or non-restoration of the EW symmetry above the TeV scale in the N2HDM (our approach could be easily applied also to other multi-scalar BSM scenarios), in a fast and computationally inexpensive way.\footnote{It is certainly much less expensive than a fully-fledged numerical minimization of the 1-loop finite temperature potential in the three-dimensional field space $(\rho_{1}(T),\rho_{2}(T),\rho_{3}(T))$.} We find that in this scenario a part of the parameter space leads to EW symmetry non-restoration at high temperatures.
It is also interesting to note that in previous studies of EW symmetry non-restoration (see
e.g.~\cite{Meade:2018saz,Baldes:2018nel}), the daisy resummation terms tend to restore the EW symmetry at high $T$, partially counterbalancing the symmetry non-restoration effect of $c_{h} < 0$ (where $c_h$ denotes the corresponding
coefficient $c_{i}$ for the case of the SM-like Higgs boson). In our study of the N2HDM we observe the opposite behavior, as the symmetry non-restoration at high temperature is driven 
by contributions from the resummation of daisy diagrams\footnote{The same behaviour has been reported very recently in~\cite{Bai:2021hfb}.} that enter the computation of the curvature with an
overall minus sign. They can prevent the restoration of the 
EW symmetry without the occurrence of negative
values of the quartic couplings, which are often in contradiction with bounded-from-below constraints. Since the daisy contributions to \refeqs{coeff_1}--\eqref{coeff_3} depend on a higher power of the scalar quartic couplings $\lambda_{j}$ than the coefficients $c_i$, the potential non-restoration behavior for the N2HDM  
arises for the case where (some of) the 
scalar couplings $\lambda_{j}$ are of \order{1}. 
Accordingly, for all benchmark points discussed in this section
we have checked the RGE running (as discussed in section~\ref{sec:rges}), ensuring that the quartic couplings satisfy the general perturbativity condition $\left|\lambda_{i}^{\overline{\rm MS}}(\mu)\right|<4\pi$ for energy scales into the tens of TeV. A detailed discussion on this issue is deferred to the following section.

Finally, we bear in mind that the study in this section has been based on the high $T$ 
expansion,\footnote{The use of only the leading $T^2$ terms guarantees here the gauge independence of the effective potential.}
and as such the $T$ dependence in Figure~\ref{c_ii_analytical} 
is expected to be fully controlled only in the high $T$ limit. 
A more detailed analysis of the intermediate $T$ 
regime should be based on the full one-loop finite-$T$ effective potential, which can only be computed numerically. 
This issue is addressed in the next section.

\subsection{Numerical Analysis}
\label{Section_benchmarks}

As discussed above, while the analytical approach developed in the previous section allows the determination of the fate of the EW symmetry at temperatures far above the TeV scale, the details of the temperature evolution from the vicinity of the EW scale upwards need to be explored numerically, since our analytical approach relies on the high-$T$ expansion of the effective potential. 
For the numerical computations, we have implemented the full one-loop effective potential given in~\refeq{1loopeffectivepotential} together with the resummed daisy contributions following the Arnold-Espinosa method, given by~\refeq{Daisyresummation}, using \texttt{CosmoTransitions}~\cite{Wainwright:2011kj} to analyse its phase/vacuum structure 
as a function of $T$. The thermal functions have been calculated using a cubic spline approximation to the exact functions given in Eq.~\eqref{thermalfunctions}. 

\begin{figure}
\centering
\includegraphics[width=0.32\textwidth]{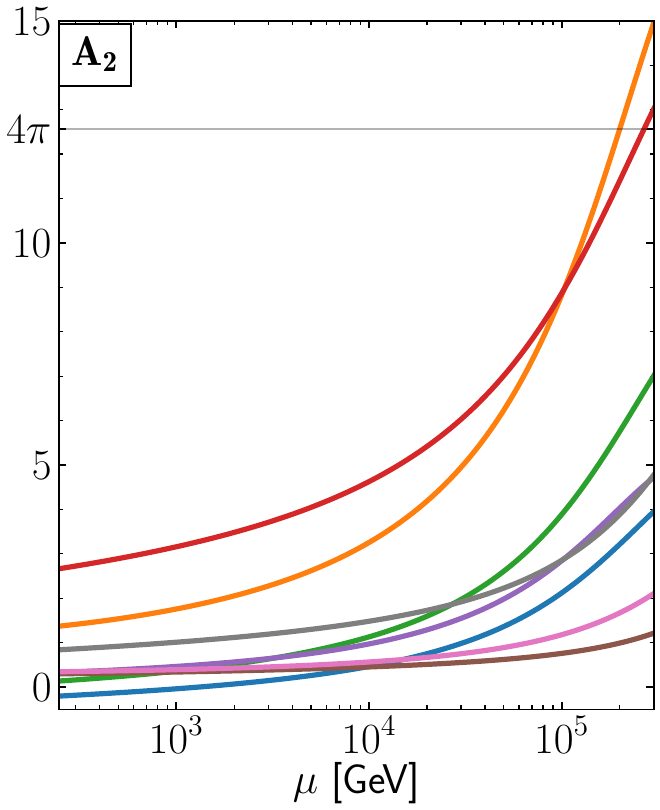}
\includegraphics[width=0.32\textwidth]{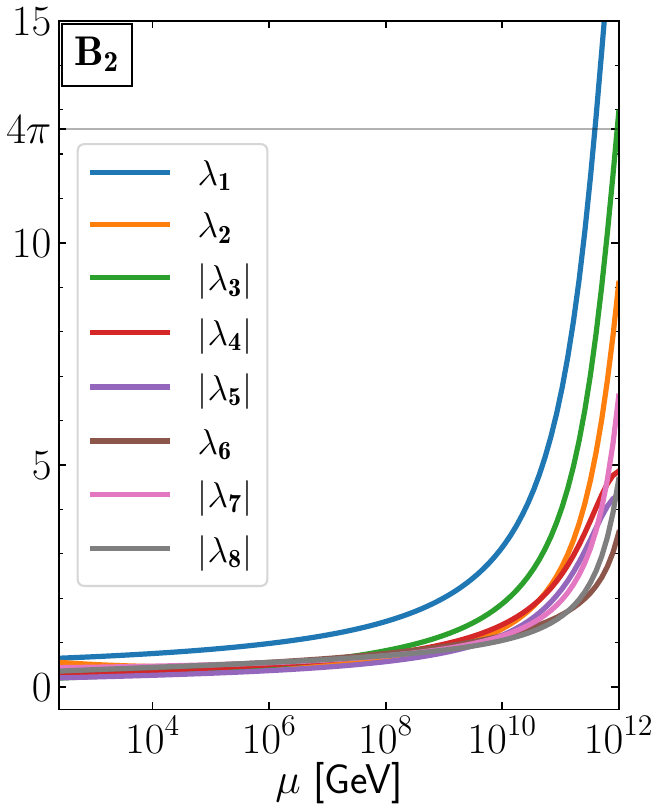}
\includegraphics[width=0.32\textwidth]{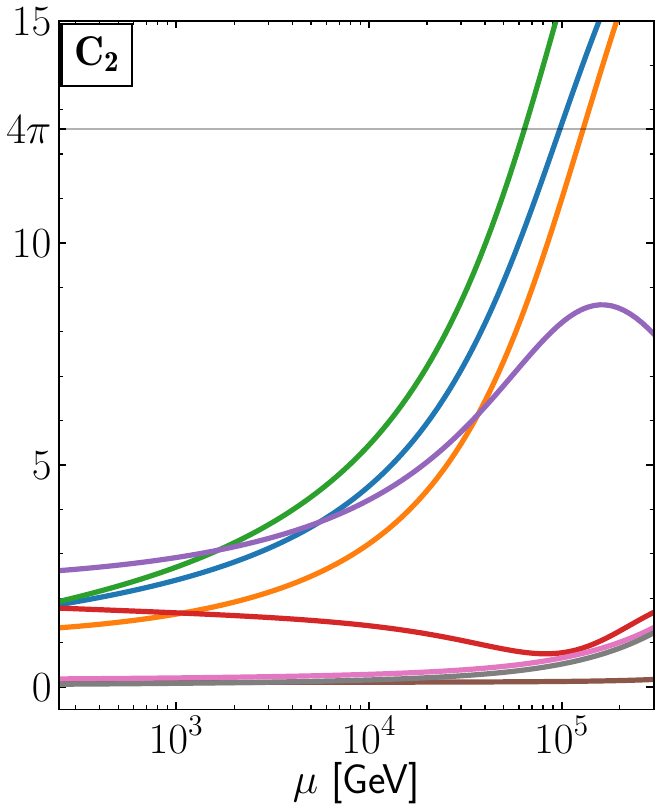}
\caption{\small Dependence of the quartic couplings $\lambda_{i}^{\overline{\rm MS}}$ on
the energy scale $\mu$ for the benchmark points A$_{2}$ (left),
B$_{2}$ (center) and C$_{2}$ (right). The gray
line indicates the perturbativity bound of $4\pi$.}
\label{pert_benchmarks}
\end{figure}

In order to ensure the validity of our numerical analysis, we first have to verify that the values of the quartic couplings $\lambda_i^{\overline{\rm MS}}$ are well within the perturbative regime over the whole temperature range that has been investigated. 
As explained in section~\ref{sec:rges}, we perform the RGE evolution of the model parameters at the two-loop level, and the energy scale $\mu$ has been varied far beyond the maximum temperature that is relevant to our analysis. To illustrate this we show in Figure~\ref{pert_benchmarks} the $\overline{\mathrm{MS}}$ values of the quartic couplings, $\lambda_i^{\overline{\rm MS}}$,
as a function of the scale $\mu$ for benchmarks A$_{2}$ (left), B$_{2}$ (middle) and C$_{2}$ (right) from Table~\ref{tab:input_parameters_benchmark_physical_2}. 
For all three benchmarks, the various $\lambda_i^{\overline{\rm MS}}$ remain perturbative up to at least $\mu = 50 \tev$, and the overall change of the couplings with the energy scale is mild below $10$ TeV due to the small coupling values at the initial scale $\mu_0 = v$. Similar results are obtained for all our benchmarks from Table~\ref{tab:input_parameters_benchmark_physical_2}.

\begin{figure}
\centering
\includegraphics[width=0.45\textwidth]{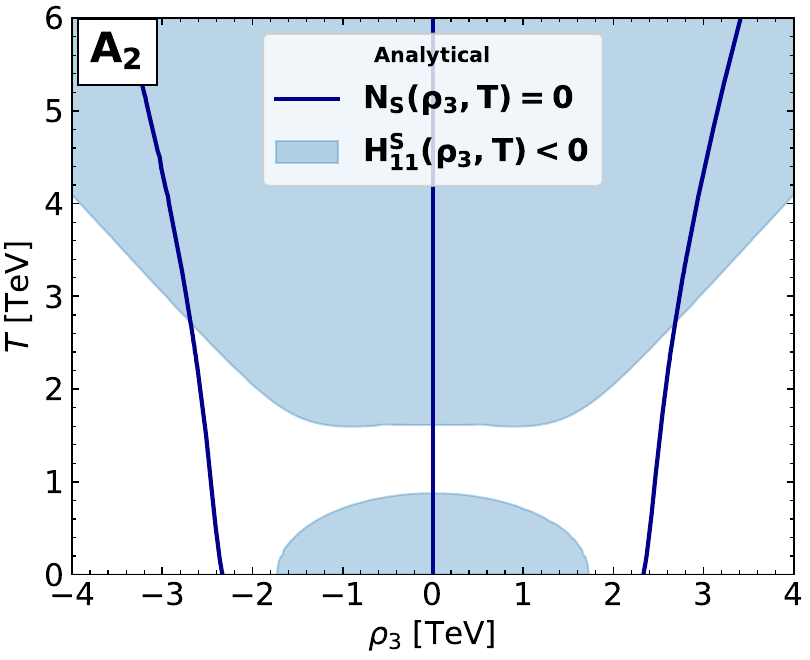}
\hspace{0.005\textwidth}
\includegraphics[width=0.45\textwidth]{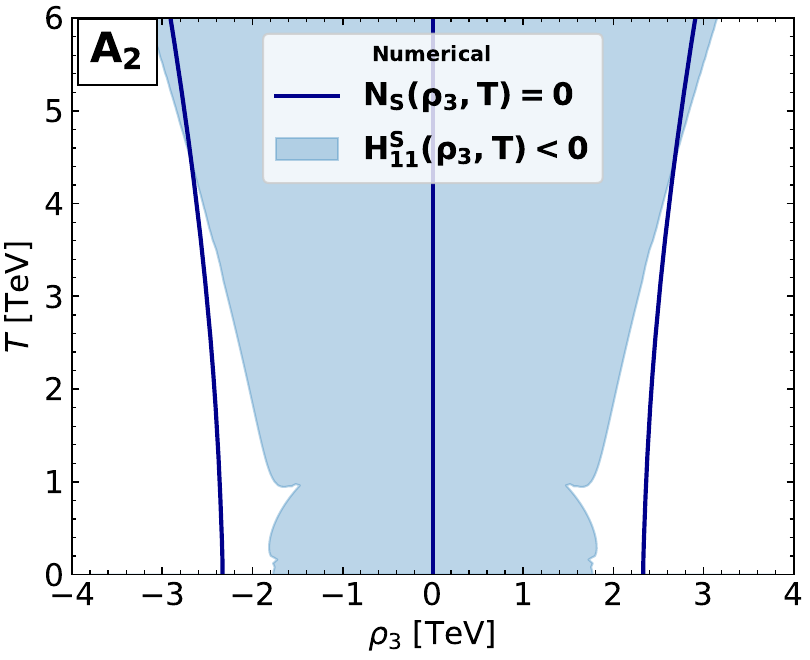}
\vspace{0.005\textwidth}
\includegraphics[width=0.45\textwidth]{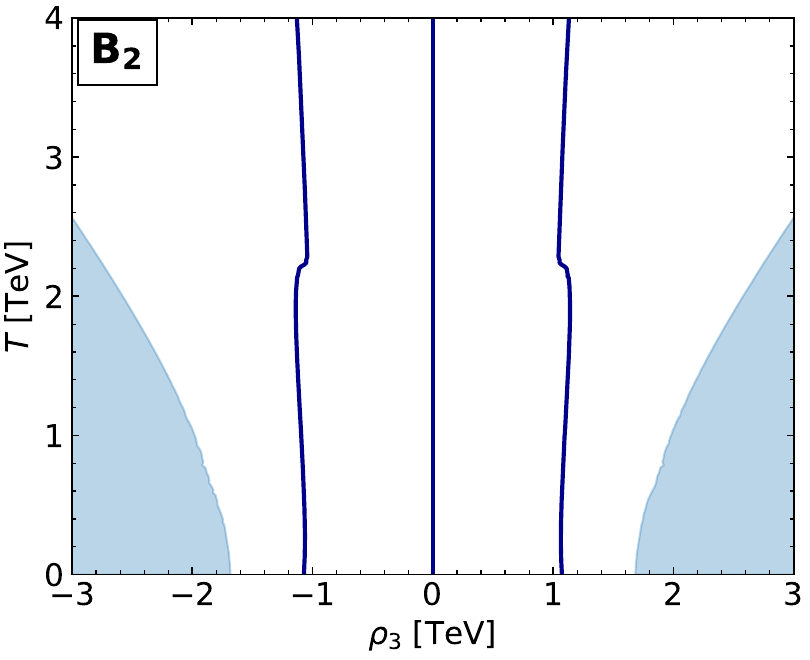}
\hspace{0.005\textwidth}
\includegraphics[width=0.45\textwidth]{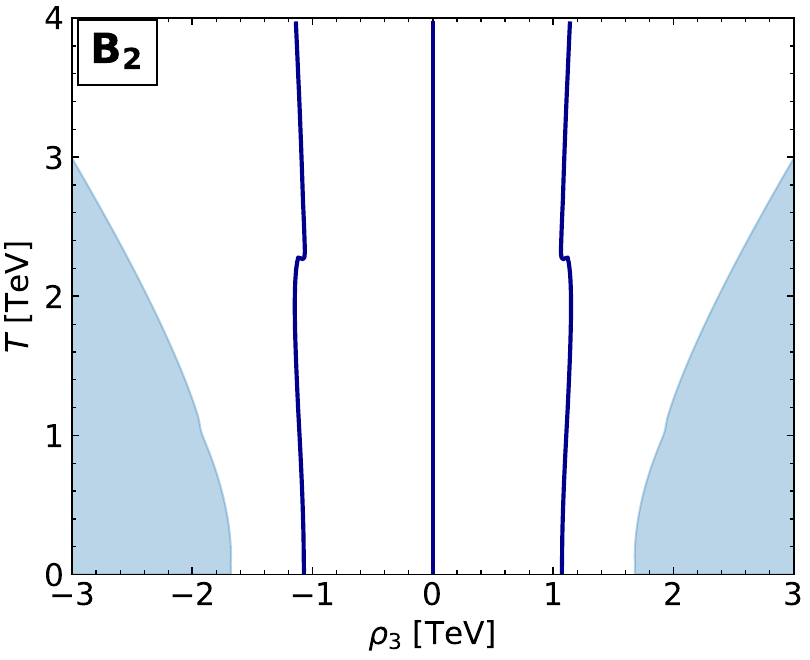}
\vspace{0.005\textwidth}
\includegraphics[width=0.45\textwidth]{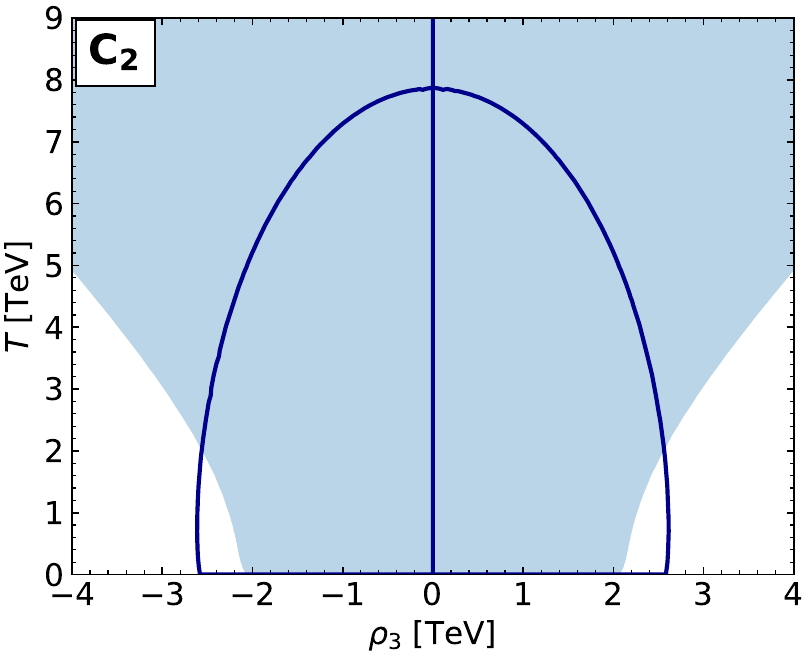}
\hspace{0.005\textwidth}
\includegraphics[width=0.45\textwidth]{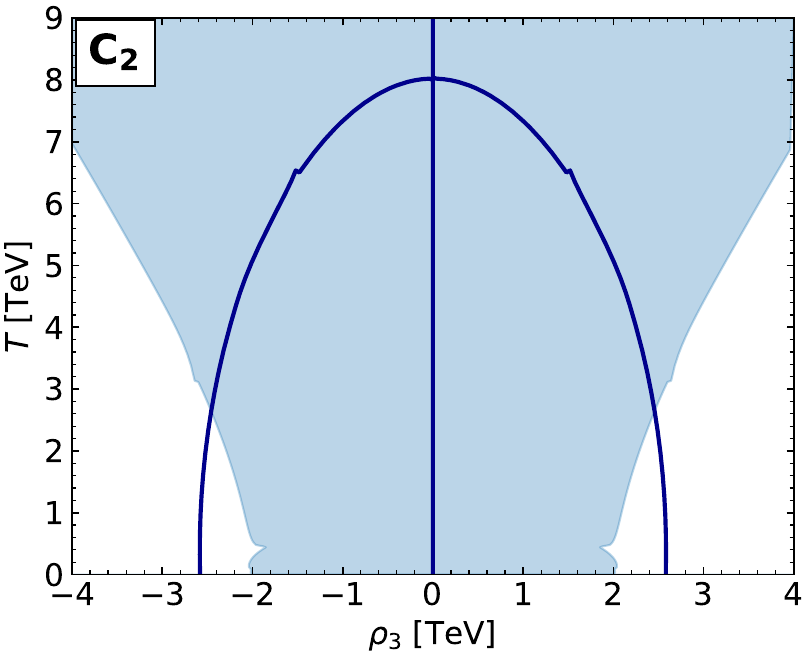}
\caption{\small Comparison of the ($\rho_{3}$--$T$) plane obtained analytically (left) and numerically (right) for the benchmark scenarios A$_2$ (top), B$_2$ (middle) and C$_2$ (bottom) from Table~\ref{tab:input_parameters_benchmark_physical_2}. The color code is the same as in Figure~\ref{c_ii_analytical}.
}
\label{c_ii_analytical_set2}
\end{figure}

We present a first comparison between our analytical and numerical analyses in Figure~\ref{c_ii_analytical_set2}, which shows (as Figure~\ref{c_ii_analytical}) the behavior of the effective potential along the $(0,\, 0,\, \rho_3)$ field space direction and in dependence of $T$, here for the benchmarks A$_{2}$ (top), B$_{2}$ (middle) and C$_{2}$ (bottom).
The predictions based on the high-$T$ approximation used in our analytical approach are shown on the left-hand side, whereas the numerical results based on the full one-loop potential are shown on the right-hand side. As in Figure~\ref{c_ii_analytical}, the dark-blue lines correspond to $N_S(\rho_{3}, T) = 0$, and the region in which $H_{11}^{S}(\rho_{3},T)<0$ is shown in light-blue. In the case of benchmark A$_2$ (upper row of Figure~\ref{c_ii_analytical_set2}), while both the numerical and analytical approaches show the non-restoration of the EW symmetry at high $T$, the shape of the $H_{11}^{S}(\rho_{3},T)<0$ region differs between the two approaches. This difference is due to the inaccuracy of the analytical treatment
in field space points in which the field values
are comparable in size to the temperature. Since $c_{33}<0$ (see Table~\ref{tab:input_parameters_benchmark_physical_2}), $|v_{S}(T)|$ grows with temperature at high $T$, 
and the scalars whose masses receive a large contribution from the singlet vev (note that $\lambda_{6,7,8}$ are sizeable for this benchmark) can therefore affect the convergence of the high-$T$ expansion.
Here it should also be noted that the derivatives
of the $J^\pm$ functions have a slower convergence towards
the corresponding high-$T$ expansions than the functions
themselves~\cite{Fowlie:2018eiu}.
At the same time, the numerical implementation of the thermal functions $J^{\pm}$ via a cubic spline introduces a small
source of uncertainty when computing numerical derivatives,\footnote{See~\cite{Fowlie:2018eiu} for a detailed discussion of the numerical issues related to the precise form of
the implementation of the thermal functions $J^\pm$ and their derivatives.}
which can also impede a better agreement between the two methods
(see also discussion below).
For benchmarks
B$_2$ and C$_2$ only minor differences arise from the uncertainties discussed above, and 
a good agreement between the analytical and numerical approaches is found.

\begin{figure}
\centering
\includegraphics[width=0.32\textwidth]{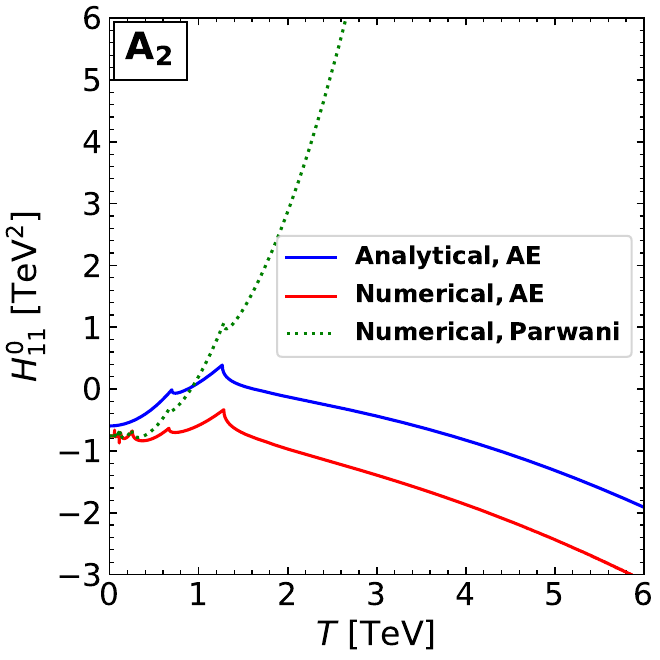}
\hspace{0.002\textwidth}
\includegraphics[width=0.32\textwidth]{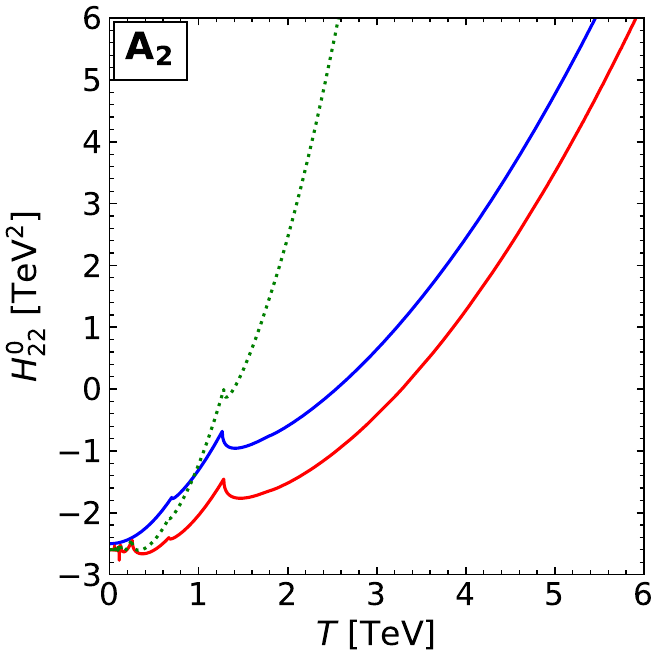}
\hspace{0.002\textwidth}
\includegraphics[width=0.32\textwidth]{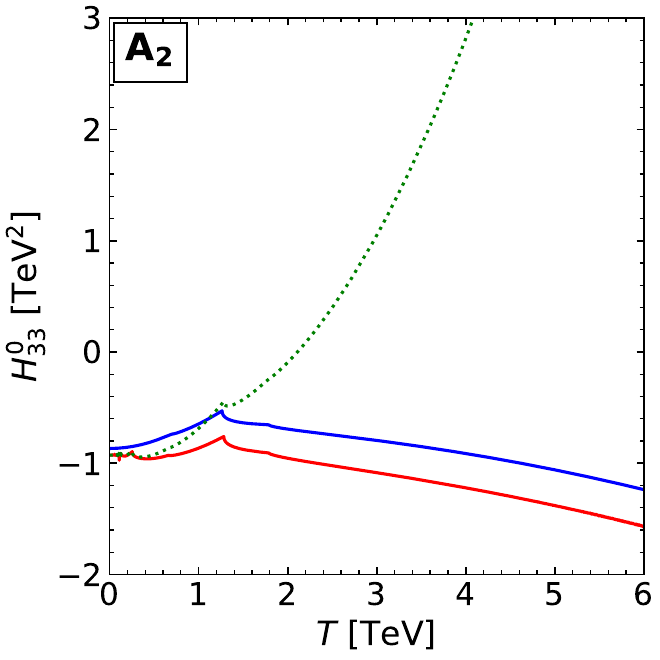}
\centering
\includegraphics[width=0.32\textwidth]{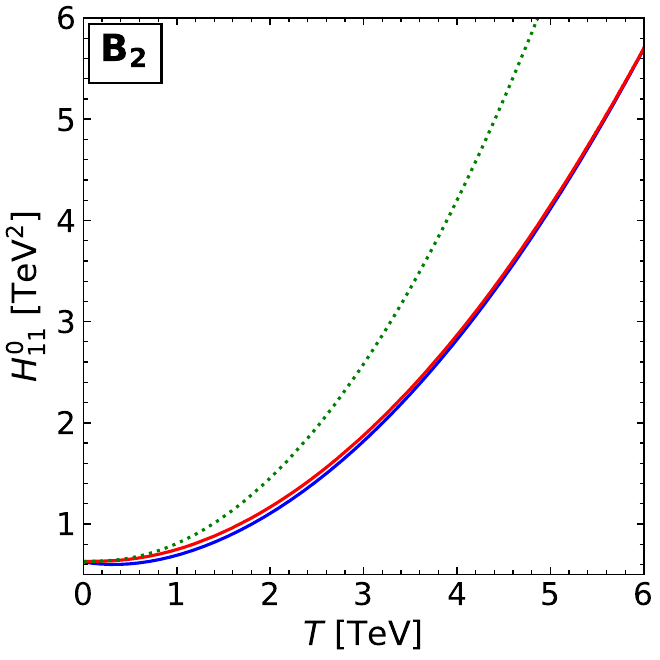}
\hspace{0.002\textwidth}
\includegraphics[width=0.32\textwidth]{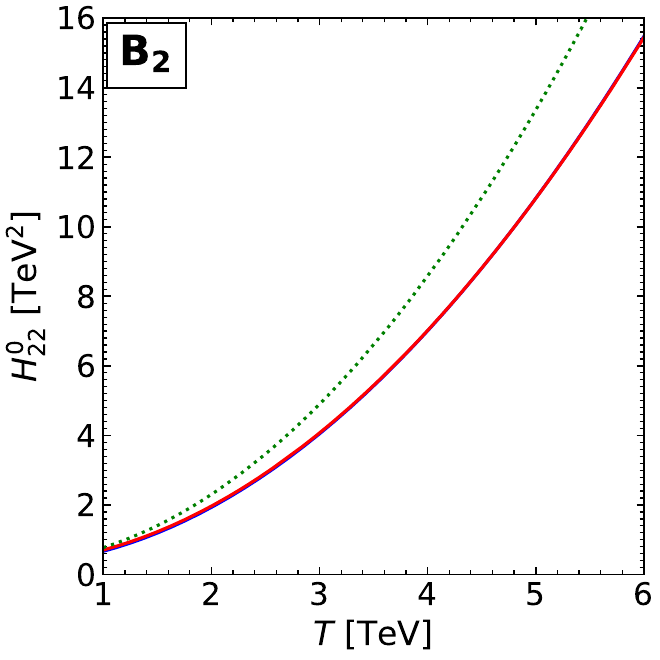}
\hspace{0.002\textwidth}
\includegraphics[width=0.32\textwidth]{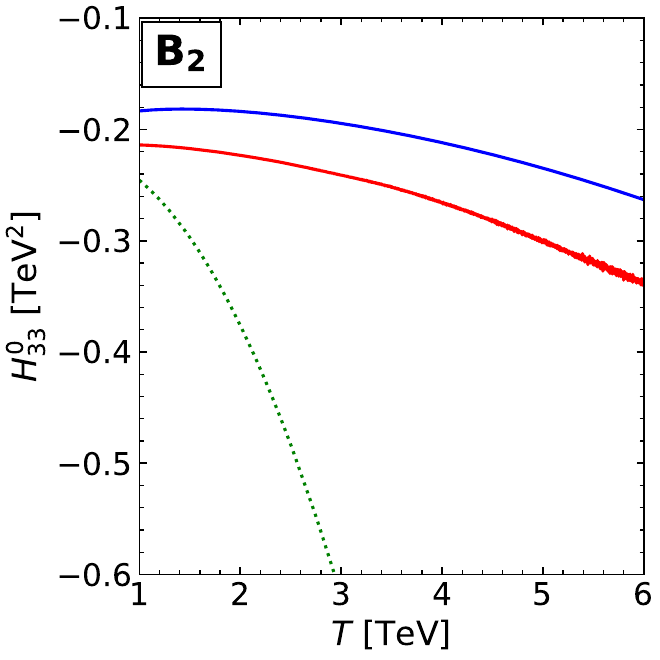}
\includegraphics[width=0.32\textwidth]{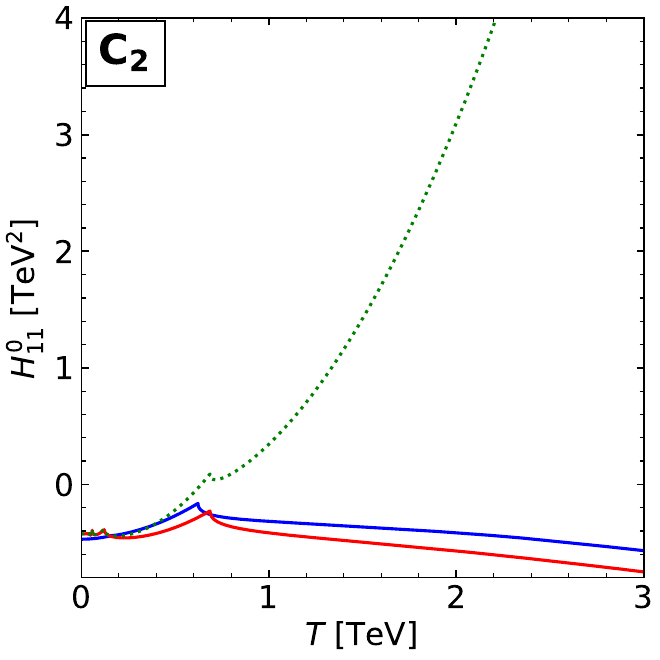}
\hspace{0.002\textwidth}
\includegraphics[width=0.32\textwidth]{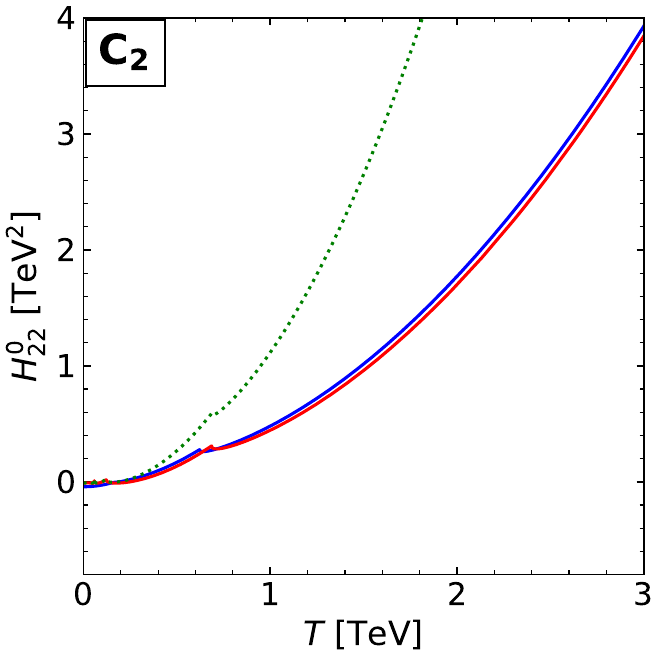}
\hspace{0.002\textwidth}
\includegraphics[width=0.32\textwidth]{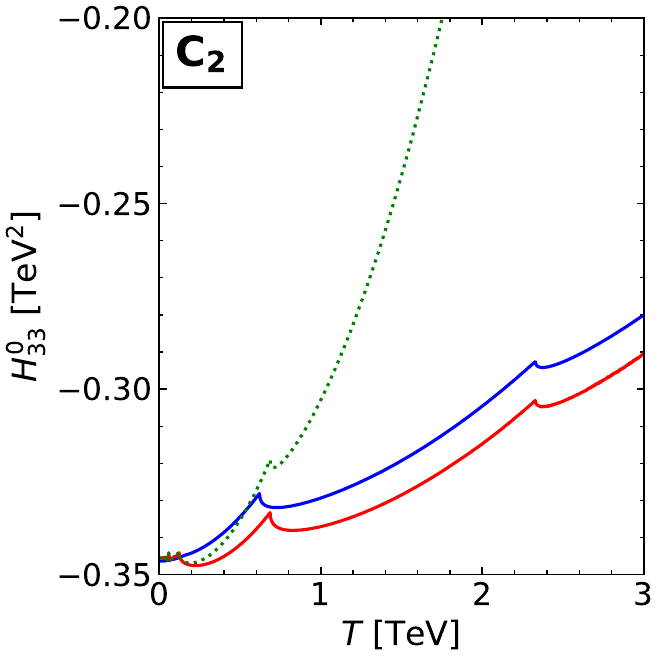}
\caption{\small Second derivatives of the effective potential at the origin of field space, $H_{ii}^{0}$, as a function of temperature for the benchmarks $\text{A}_2$ (upper row), $\text{B}_2$ (middle row) and $\text{C}_2$ (lower row) using the
analytical (within the Arnold-Espinosa -- AE -- approach) high-$T$ approximation (blue) 
and evaluated numerically 
using the Arnold-Espinosa  approach (red) and the Parwani approach (dashed
green).}
\label{hessian_comparison}
\end{figure}

In order to better understand the differences between the analytical and the numerical approach, in Figure~\ref{hessian_comparison} we show for the three benchmark points used above
the curvatures at the origin of field space, $H^0_{ii}$, as a function of temperature. In the numerical analysis, we compare the
Arnold-Espinosa (solid-red lines) and Parwani (dotted-green lines) approaches.
The values of $H^0_{ii}$ computed analytically using the high-$T$ expansion (according to the Arnold-Espinosa approach, see the discussion in section~\ref{sec:sinnlos}) are also depicted (solid-blue lines). 
One can observe that the Parwani method agrees with the Arnold-Espinosa numerical result only for very small~$T$ and thus can lead to a different prediction for the restoration of both the EW and the \Ztp symmetries.\footnote{The Parwani method inconsistently mixes loop contributions of different orders in the couplings and the temperature 
(see e.g.~\cite{Laine:2017hdk,Kainulainen:2019kyp}),
 such that thermal effects are enhanced as compared to the Arnold-Espinosa approach.
Using the latter, the cancellation between
logarithmic contributions between $V_{\rm CW}$ and
$V_T$ are ensured in the high-$T$ limit.} 
This is indicated by the fact that in several cases 
the result for $H^0_{ii}$ grows with temperature using the Parwani implementation, while the result for $H^0_{ii}$ obtained with the Arnold-Espinosa method decreases with temperature (see, for instance, the upper left and upper right plot of \reffi{hessian_comparison}). Thus, these two methods employed in the literature to resum the contributions from daisy diagrams may yield qualitatively different behaviors at high temperature, as was already pointed out in section~\ref{sec:sinnlos}  (see also \refeq{higher_orders_expansion}).

We also see from Figure~\ref{hessian_comparison} that our $H^0_{ii}$ analytical computation 
agrees well with
the Arnold-Espinosa numerical result.
Only a minor difference is present between the numerical result
and the analytical one (both rely on the Arnold-Espinosa
method). This difference is mainly driven by the fact
that the $\mathcal{O}(y^3)$ and, in particular, the logarithmic
pieces 
of the expansions of the $J^\pm$
functions are not taken into
account in the analytical treatment, whereas they are implicitly
contained in the numerical result, for which the thermal
functions are implemented as an interpolation to the
exact form of the integrals. As a result of this interpolation,
also the numerical predictions for the second derivatives
suffer from an uncertainty at large temperatures, due
to the fact that the tiny variations of the functional argument
$y$ cannot be resolved.\footnote{We used a step size
of $y_{\rm step} = 0.01$ for the  
cubic
spline interpolation of $J^\pm(y)$.}
In combination, the uncertainties of both methods
give rise to the small offsets between the blue and
the red curves that are visible in
\reffi{hessian_comparison}. These differences however do not affect 
our results for the EW symmetry restoration behavior, for which the predictions obtained using the numerical and the analytical method within the Arnold-Espinosa approach are very similar
In view of the discussion in the previous section, this adds support
to the reliability of our analytical analysis for predicting the fate of the EW symmetry at high temperature in the N2HDM~\cite{Kainulainen:2019kyp}. In particular, if $c_{33} > 0$ or if $c_{33} < 0$ and $v_{S}(T)^2/T^2$ is sufficiently small at high $T$ (see Eq.~\eqref{H11S_limit}), the coefficient $c_{11}$ controls the stability of the effective  potential along the doublet field directions at high $T$.

An analogous approach can be applied to the study of the fate of the discrete \Ztp symmetry as a function of temperature. In this case, however, we can only make definite analytical statements regarding the restoration of the \Ztp symmetry in a handful of scenarios: if both $c_{11}$ and $c_{33}$ are positive, the \Ztp symmetry is restored at high $T$, while if $c_{33} < 0$ and $c_{11}^S > 0$ then the \Ztp symmetry is broken at high $T$. The investigation of other scenarios would require a numerical analysis of the finite temperature effective potential for $\left\langle \Phi_{1}
\right\rangle \neq 0$, $\left\langle \Phi_{2} \right\rangle \neq 0$, which can be in principle performed with \texttt{CosmoTransitions}, but which we do not pursue further in this work.  

\medskip
The cosmological consequences of the possible non-restoration of EW and/or \Ztp symmetries are several:
domain wall problems in cosmology are associated with the existence of multiple vacua in theories with spontaneous breaking of discrete symmetries. However, N2HDM scenarios in which the \Ztp symmetry is never restored would trivially avoid the formation of domain walls, i.e.\ eliminating the domain-wall problem. 
On the other hand, an un-restored EW symmetry at high temperatures would lead to a very strong suppression of the baryon-number-violating sphaleron transitions at those temperatures, possibly hindering baryogenesis/leptogenesis mechanisms relying on 
sphalerons. 
Yet, we stress that high-$T$ EW symmetry non-restoration is not incompatible with having $H^0_{11} > 0$ at intermediate temperatures, as shown e.g.\ in Figure~\ref{hessian_comparison}, top-left (see also Figure~\ref{c_ii_analytical_set2}, top-left).
This means that the EW phase transition could take place also in such scenarios. We will explore this possibility in more detail in the next section.

\subsection{The EW phase transition and symmetry non-restoration}
\label{SNR_FOEWPT}

In this section we explore the possibility of a FOEWPT in the N2HDM and discuss its connection to the possible non-restoration of the EW symmetry at high $T$. 
The simultaneous occurrence of both phenomena requires a temporary restoration of the EW symmetry, together with its breaking at higher temperatures.

\begin{table}
\centering
{\renewcommand{\arraystretch}{1.4}
\footnotesize
\begin{tabular}{ccccccccccc}
$m_{h_{a}}$  & $m_{h_{b}}$  & $m_{h_{c}}$ & $m_{A}$ &
$m_{H^{\pm}}$  & $\text{tan}\beta$ &
$C^{2}_{h_{a}t\bar{t}}$ & $C^{2}_{h_{a}VV}$  & $R_{b3}$  & $m_{12}^{2}$ & $v_{S}$ \\
\hline
\hline
$125.09$  & $[30,1000]$  & $400$ & $650$ &
$650$  & $2$ &
$1$ & $1$  & $[-1,1]$  & $65000$ & $[1,1000]$
\end{tabular}
}
\caption{\small Set of input parameters for our \texttt{ScannerS} scan. All the input parameters remain fixed except for the mass of one of the CP-even Higgs bosons $m_{h_{b}}$, its singlet component $R_{b3}$ (with  $\Sigma_{h_{b}}=|R_{b3}|^2$) and the singlet vev at $T=0$, $v_{S}$.}
\label{tab:input_parameters_Scan}
\end{table}

In order to investigate the parameter region of the N2HDM possibly realising a FOEWPT, we start by discussing the region of the 2HDM featuring a FOEWPT, and analyse in the next step how the presence of the singlet field $\Phi_S$ in the N2HDM affects this picture. In the type-II 2HDM the region of parameter space giving rise to a FOEWPT is quite constrained: it generally correlates with the  existence of sizeable quartic couplings among $\lambda_{3,4,5}$ in the 2HDM scalar potential. Since the mass splittings between the 2HDM scalars are also controlled by such couplings,  at least one of the additional 2HDM scalars (apart from the SM-like Higgs boson at about $125\gev$) must be significantly lighter or heavier than the overall mass scale $M=\sqrt{m_{12}^{2}/(\sin\beta\cos\beta)}$ of the second Higgs doublet~\cite{Dorsch:2017nza}.
Therefore, in general a FOEWPT in the 2HDM relies on a hierarchical spectrum with a considerable mass 
splitting between the pseudoscalar $A$ and the heaviest CP-even Higgs boson $H$~\cite{Dorsch:2014qja, Dorsch:2013wja, Dorsch:2016nrg}. 
In the type-II 2HDM $B$-physics observables push the mass of the charged scalar to $m_{H^{\pm}}>590\gev$~\cite{Haller:2018nnx}. In combination with EWPO constraints, this results in a
most obvious possibility for the  
realization of a FOEWPT in a type-II 2HDM consisting on a hierarchical spectrum with
$m_A \approx m_{H^\pm} \gtrsim 600\gev$ and a substantially lighter scalar state~$H$.\footnote{The opposite case with
$m_H \approx m_{H^\pm} \gg m_A$ is much less favorable for
FOEWPTs, as it requires almost exact alignment in order
to decouple the heavy $H$ from the phase transition
dynamics~\cite{Dorsch:2013wja,Dorsch:2014qja,Dorsch:2016nrg}.}

Based on the above considerations for the 2HDM, we generated a total of 2000 N2HDM benchmark points with \texttt{ScannerS} using values for the free parameters as shown in Table~\ref{tab:input_parameters_Scan}, with a flat prior for the parameters $m_{h_b}$, $R_{b3}$ and $v_S$ that have been varied.
All the benchmarks fulfill the theoretical and experimental constraints
described in section~\ref{constraints}. The chosen mass gap between $m_{h_c}$ and $m_A$ increases the possibility of a FOEWPT in analogy to the 2HDM case. We also focused on the alignment limit with $C^{2}_{h_{a}tt}=C^{2}_{h_{a}VV}=1$, such that only $h_b$ and $h_c$ can have a non-zero singlet admixture
(hence $R_{a3}=0$). The parameters varied in our scan correspond to those related to the presence of the singlet in the N2HDM Higgs sector, i.e.~the mass of a third CP-even Higgs boson $m_{h_{b}}$, its singlet component $R_{b3}$ and the value of the singlet vev $v_{S}$.

Using \texttt{CosmoTransitions}, we have numerically analyzed the thermal history of each scan point within the temperature regime $T = [0, 
600 \gev]$, which covers the region relevant for the possible presence of a FOEWPT. At $T=0$, we observe that each point has a global EW minimum of the kind $(v_1,v_2,v_S)$ and a false minimum of the kind $(0,0,\tilde{v}_S)$, separated by a potential barrier generated already at $T = 0$ by $V_{\rm CW}$.
We find that 542 points out of the 2000 initial \texttt{ScannerS} benchmarks feature a FOEWPT. 
Most of the scan points satisfy the general perturbativity conditions $|\lambda_{i}^{\overline{\rm MS}}|<4\pi$ up to energy scales larger than ${\mu =2 \tev}$. At the same time, we find that our analytical analysis to ascertain the fate of the EW symmetry at high temperature is, for a large set of the scan sample, already
applicable for $T \lesssim 1$ TeV, given that the mass scale of the doublet field bilinears is close to the EW scale for most of the benchmarks (e.g.~the mass scale of the non-SM Higgs doublet is $M \approx 400\gev$).

\begin{figure}
\centering
\includegraphics[width=0.328\textwidth]{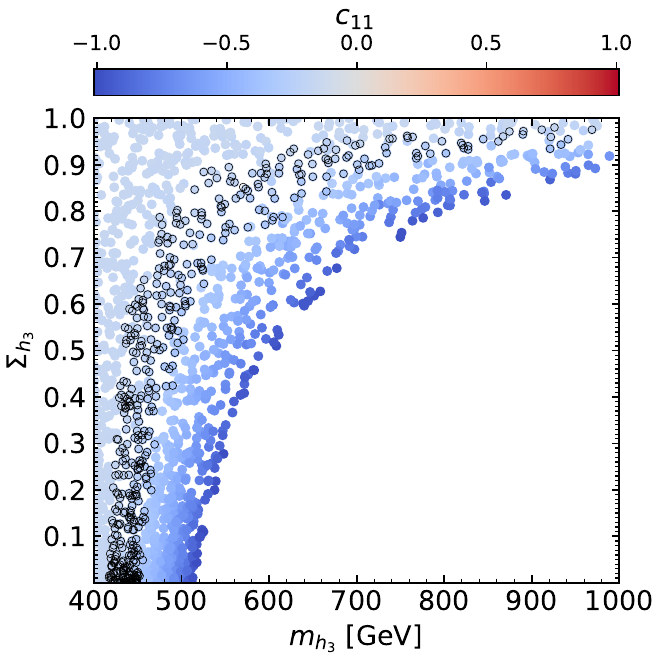}
\hspace{-0.01\textwidth}
\includegraphics[width=0.328\textwidth]{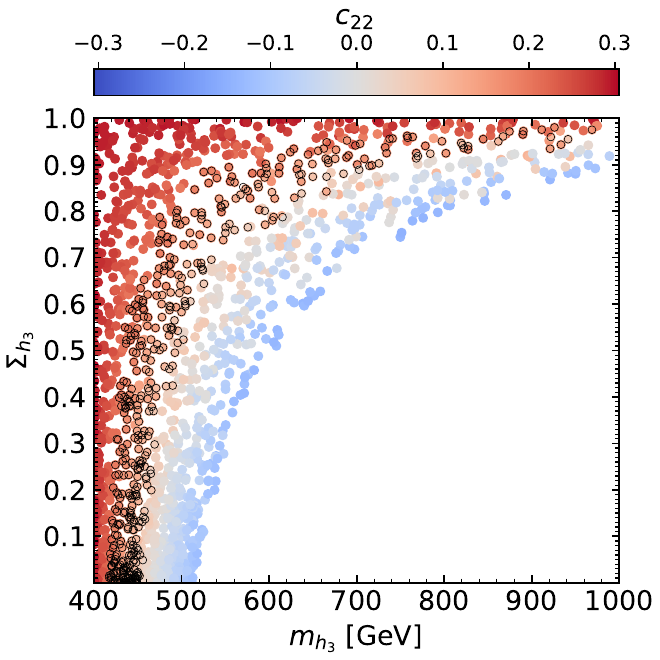}
\hspace{-0.01\textwidth}
\includegraphics[width=0.328\textwidth]{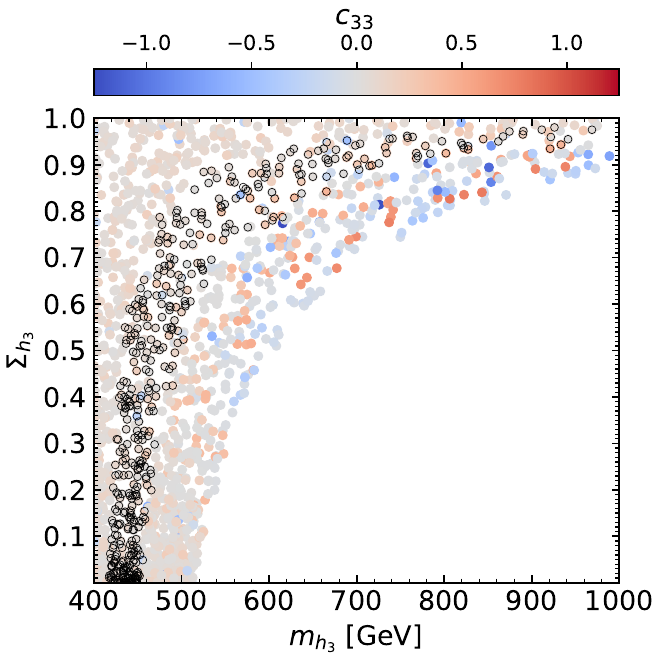}
\caption{\small Singlet component $\Sigma_{h_{3}}$ of the third CP-even Higgs boson in
dependence of its mass $m_{h_{3}}$. The color code indicates the value of the
coefficients
$c_{11}$ (left), $c_{22}$ (middle) and $c_{33}$
(right). Points with a black circular edge feature a FOEWPT.
}
\label{SNR_Jamaica}
\end{figure}

In Figure~\ref{SNR_Jamaica} we show the results of our \texttt{ScannerS} parameter scan in the  ($m_{h_3}$--$\Sigma_{h_3}$) plane, i.e.~the heaviest CP-even scalar mass vs.\ its singlet component, with the color code indicating the value of the coefficients $c_{11}$ (left),
$c_{22}$ (center) and $c_{33}$ (right).
The absence of points in the lower right region 
is due to the perturbative unitarity constraints.
The points that feature a FOEWPT are highlighted with a black circular edge. We see that none of the points of our scan features the restoration of both the EW and \Ztp symmetries at high temperature, given that $c_{11}<0$ for all points. 
The coefficient $c_{22}$, even though positive for most benchmarks in the scan, reaches negative values for a fraction of the points. However, there is no point with $c_{22} < 0$ and $c_{11} > 0$ due to the positive $Y_t$ contribution to $c_{22}$, which confirms our expectation that $c_{22}$ provides no relevant information for the fate of the EW symmetry at high $T$ (see the discussion in section~\ref{SNR_framework}).
The coefficient $c_{33}$, which is related to the possible restoration of the $\mathbb{Z}_{2}'$
symmetry, obtains values in our scan that range from $\approx -1.25$ to $\approx 0.8$. 
Figure~\ref{SNR_Jamaica} highlights that it is perfectly possible to have a FOEWPT together with an unrestored EW symmetry at higher temperatures within the N2HDM. Actually, all the scan points featuring a FOEWPT have $c_{11} < 0$ (note however that as discussed above $c_{11} < 0$ is only a sufficient condition for EW symmetry non-restoration at high $T$ if $c_{33} > 0$). The connection between both phenomena lies in the fact that the sizeable scalar quartic couplings which contribute to making the EW phase transition strongly first-order in the (N)2HDM may also
contribute to rendering the coefficients $c_{ii}$ negative, see Eqs.~\eqref{coeff_1}--\eqref{coeff_3}.

\begin{table}
\centering
{\renewcommand{\arraystretch}{1.4}
\footnotesize
\begin{tabular}{c||cccccccccc||cc}
& $m_{h_{1}}$  & $m_{h_{2}}$  & $m_{h_{3}}$ & $m_{A,H^\pm}$  &
$t_{\beta}$ & $C_{h_{1}tt/VV}$ &
$\text{sgn}\left(R_{13}\right)$  & $R_{23}$  & $m_{12}^2$ & $v_{S}$ & $c_{11}$ & $c_{33}$ \\
\hline
\hline
E$_1$ & $125.09$  & $400$  & $517.29$ & $650$ & $2$ & $1$ & $-1$ &
$0.98$  & $255^2$ & $746.79$ & $-0.99$ & $0.007$ \\
\hline
E$_2$ & $125.09$  & $400$  & $487.29$ & $650$ & $2$ & $1$ & $-1$ &
$0.64$  & $255^2$ & $559.31$ & $-0.24$ & $0.03$ \\
\hline
E$_3$ & $125.09$  & $400$  & $550.25$ & $650$ & $2$ & $1$ & $-1$ &
$0.48$  & $255^2$ & $544.63$ & $-0.25$ & $0.03$ \\
\end{tabular}
}
\caption{\small Type-II N2HDM benchmarks corresponding to the scan described by the
input parameters in Table~\ref{tab:input_parameters_Scan}.
Point E$_1$ is a benchmark point for which
the EW symmetry is never restored up to
the maximum temperature analysed, $T_{\text{max}}$, E$_2$ features a
FOEWPT, and E$_3$ features a Universe trapped in a false
vacuum at zero temperature.
Also shown are the values of $c_{11}$ and $c_{33}$
for 
the three benchmark points.
$c_{22}$ is positive for the three scenarios.
The parameters $m_{h_i}$, $m_A$, $ m_{H^{\pm}}$, $m_{12}$ and $v_S$ are given in GeV.}
\label{tab:benchmarks_Jamaica}
\end{table}

\begin{figure}
\centering
\includegraphics[width=0.45\textwidth]{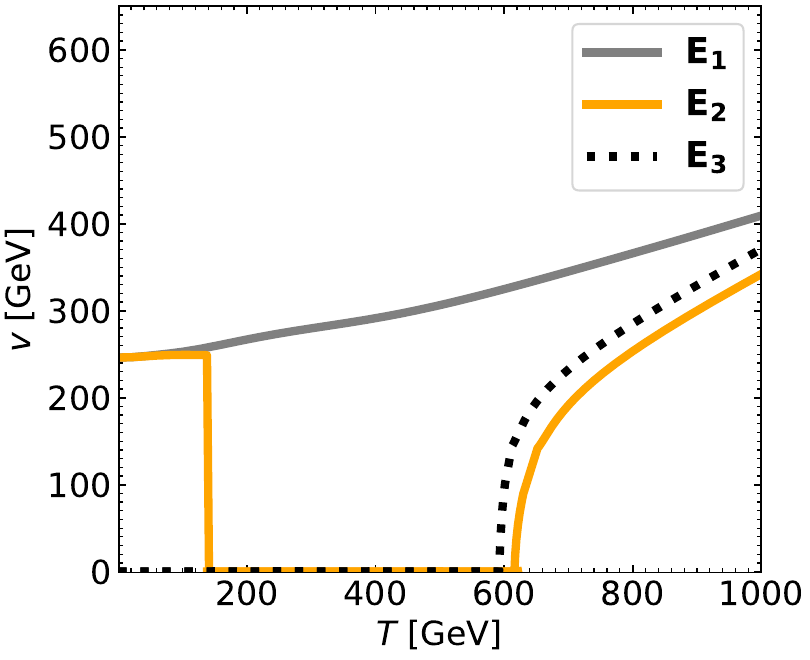}
\hspace{0.002\textwidth}
\includegraphics[width=0.45\textwidth]{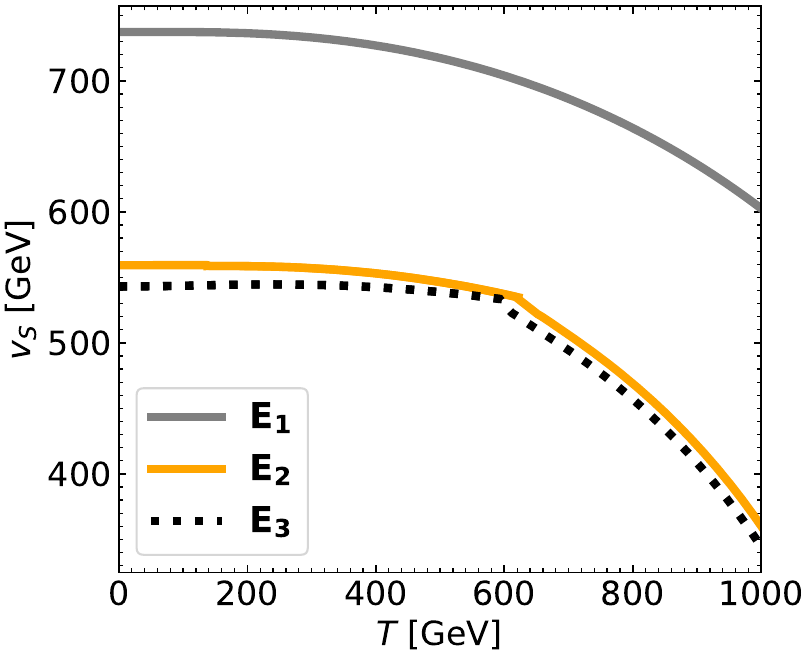}
\caption{\small Electroweak vev $v$ (left) and singlet vev $v_{S}$ (right) in dependence of the temperature for the
benchmarks E$_1$ (solid gray), E$_2$ (orange) and E$_3$ (dashed black). These curves were computed using our numerical
implementation of the full 1-loop potential in \texttt{CosmoTransitions}. }
\label{VEVS_Jamaica_T}
\end{figure}

In the following we choose three benchmark points from our parameter scan to illustrate different EW thermal histories (we give further details on the rationale behind this choice in the next section).
The \texttt{ScannerS} input parameters of the three benchmarks E$_{1,2,3}$ are given in Table~\ref{tab:benchmarks_Jamaica},
in which we also show the values for their coefficients $c_{11}$ and $c_{33}$.
For these three benchmarks, we also use \texttt{CosmoTransitions} to numerically track the evolution of the vacuum of the system\footnote{We take into account all possible local minima at each temperature in the three-dimensional field space.} ($v(T)$, $v_{S}(T)$) from a temperature $T_{\rm max} = 1\tev$ down to $T = 0$. 
We show in Figure~\ref{VEVS_Jamaica_T} the temperature evolution of the EW vev $v(T)$ (left) and the singlet vev $v_{S}(T)$ (right) for each of the three benchmarks. For scenario E$_{1}$ the EW symmetry is never restored up to 
$T_{\rm max}$, and no EW phase transition occurs in this temperature range, in agreement with the expectation from the  values of $c_{11}$ and $c_{33}$ for this benchmark (see 
Table~\ref{tab:benchmarks_Jamaica}).
In contrast, scenario E$_{2}$ shows a FOEWPT with a nucleation temperature $T_{n}=138 \gev$, and the EW symmetry is restored in the temperature range $T \in [T_n, \,620 \gev]$. For temperatures larger than $\approx 620$ GeV the vacuum with $ v(T) = 0$ is unstable, and the EW symmetry is thus un-restored at high $T$.  
The singlet vev $v_S(T)$ starts to decrease for $T \gtrsim 600 \gev$, which suggests that at very high temperatures the Universe would have been in an EW-breaking but \Ztp-conserving vacuum configuration (this is in agreement with the values of $c_{11}$ and $c_{33}$ found for this scenario).
Finally, for benchmark scenario E$_{3}$ the EW symmetry is broken at high temperatures and becomes 
unbroken when the Universe reaches a temperature $T \approx 600 \gev$. 
However, for lower temperatures the Universe does 
not undergo another transition to the EW vacuum,\footnote{The point E$_{3}$ shows the peculiar phenomenon that the EW symmetry is broken at high $T$ 
but unbroken at $T=0$, the opposite of the commonly expected behavior.} but rather it is trapped in an EW symmetric phase down to $T = 0$, which makes this scenario unphysical. The existence of these \textit{trapped-vacuum} scenarios in the N2HDM has already been briefly discussed in section~\ref{sec:thermalhistory}, and we will explore it in more detail in section~\ref{section_trapped}.

Before moving on to the next section, we note that for all three benchmarks E$_{1,2,3}$ the value of the singlet vev $v_{S}(T)$ shows a decreasing trend for increasing temperature, as shown in Figure~\ref{VEVS_Jamaica_T}.
Yet, the \Ztp symmetry is not restored at $T_{\rm max}$ and one would have to go to larger temperatures to observe its restoration (as expected from the value of $c_{33}$ for all three benchmarks, see Table~\ref{tab:benchmarks_Jamaica}). However, the values of the quartic couplings $\lambda_{i}$ for these benchmarks are relatively large, and we find that the RGE evolution of $\lambda_{i}$ cannot be neglected for energies above 1~TeV. We therefore did not investigate here the behavior of the effective potential beyond $T_{\rm max} = 1 \tev$, which would require including the effect of this RGE evolution.

\section{Trapped metastable singlet vacua}
\label{section_trapped}

In section~\ref{sec:thermalhistory} we pointed out the phenomenon of \textit{vacuum trapping}: 
in the N2HDM the situation can occur that a scenario with a global EW minimum at $T = 0$ is in fact unphysical when the Universe is trapped in a false (singlet) vacuum. Vacuum trapping is expected to be particularly relevant in parameter regions in which FOEWPTs can occur, as for such regions several minima that are simultaneously present in the early Universe (one of them being the EW-broken phase) can co-exist down to $T=0$ (see also the analysis of the vacuum structure of the N2HDM at $T = 0$ that was performed in~\cite{Ferreira:2019iqb}).
In this section we explore in more detail the occurrence of (singlet) vacuum trapping, and discuss its relation to the N2HDM parameter space region featuring a FOEWPT.
Our analysis is divided into two different parts, which concentrate on two regions of the N2HDM parameter space with different phenomenological features: in section~\ref{jamaica} we analyse scenarios with a SM-like Higgs boson at about 125 GeV, $h_{125}$, where the singlet field mixes with the 2HDM-like
heavy CP-even scalar $H$;  
in section~\ref{mixhsmsing} we then analyse scenarios in which 
$h_{125}$ contains a singlet admixture, while the singlet field does not mix with $H$. 
In both cases we study the interplay between scenarios with trapped singlet vacua and with a FOEWPT, as well as the connection of such early Universe phenomena to the phenomenology of the N2HDM.

In the following we also emphasize the conceptual differences between the N2HDM and the (extensively studied) 2HDM regarding the EW phase transition. In the N2HDM, due to the presence of the singlet field and its associated \Ztp symmetry, there can be several phase transitions during the thermal history of a specific N2HDM scenario.
Yet, since the singlet field does not couple directly to the massive degrees of freedom of the SM (besides the Higgs sector), the breaking of the \Ztp symmetry usually takes place at higher temperatures than the
possible FOEWPT.\footnote{Only for very small values of $v_S \ll v$
a simultaneous breaking (in a single transition) of the EW and \Ztp symmetries can be realized. 
A discussion of this scenario is
left for future work.
Conversely, in our analysis with $v_S > v$
there is always a large temperature gap between the
transition into the trapped vacuum
$(0,0,\tilde{v}_S)$ and the
formation of the EW vacuum
$(v_1,v_2,v_S)$.}
Thus, the FOEWPTs that we analyze in the following are  of the type $(0,0,\tilde{v}_S(T)) \to (v_1(T),v_2(T),v_S(T))$, where $\tilde{v}_S(T)$ and $v_S(T)$ are the vevs of the singlet field in the EW-conserving \textit{false} minimum and the EW-breaking \textit{true} minimum, respectively. This type of transitions obviously does not exist in the case of the 2HDM.

\subsection{Case 1: Singlet admixture in \texorpdfstring{$H$}{Hheavy}}
\label{jamaica}

We focus here on scenarios where the singlet scalar field mixes only with the heavy CP-even 2HDM state $H$, while the SM-like Higgs boson at about $125$ GeV is unaffected by the mixing with the singlet state. We use again the N2HDM parameter scan discussed in section~\ref{SNR_FOEWPT} in connection to EW symmetry non-restoration, as defined in Table~\ref{tab:input_parameters_Scan}.
We note that the scan parameter values, and in particular the hierarchy between ${m_{h_c}}$ and $m_A = m_{H^\pm}$, have been chosen so as to explore in detail the impact of the presence of the singlet field within the N2HDM, relative to scenarios that would feature a FOEWPT in the 2HDM~\cite{Dorsch:2013wja,
Dorsch:2014qja,Basler:2016obg,Bernon:2017jgv}.

\begin{figure}[t]
\centering
\includegraphics[width=0.48\textwidth]{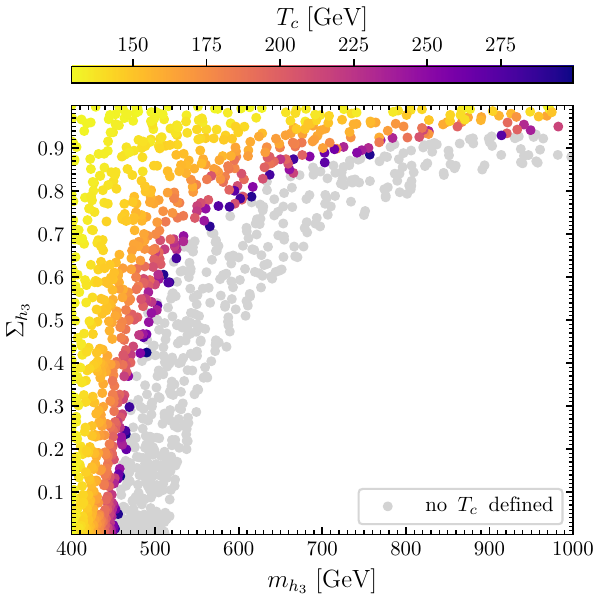}~
\includegraphics[width=0.48\textwidth]{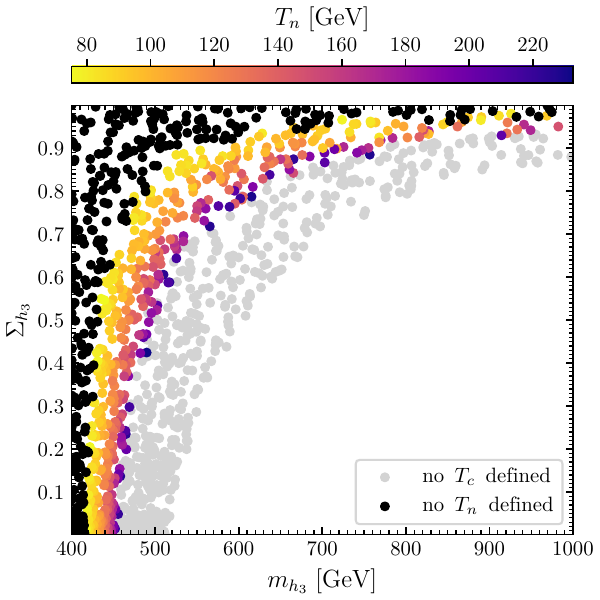}
\caption{\small Parameter points according to Table~\ref{tab:input_parameters_Scan}. The singlet
component $\Sigma_{h_3}$ of the third CP-even Higgs boson is shown in dependence of its mass $m_{h_{3}}$. The color code indicates the values for the critical temperature $T_c$ (left)
and the nucleation temperature $T_{n}$
for the points with a FOEWPT (right). Black
points do not feature a $T_{n}$, and the Universe is trapped in a false minimum. For the gray points, for which $T_{c}$ cannot be defined, the Universe is in the EW minimum already at $T_{\rm max}=600\gev$, i.e.\ the EW symmetry is
not restored within the investigated temperature range, and no FOEWPT occurs.}
\label{Jamaica_scan}
\end{figure}

In Figure~\ref{Jamaica_scan} we show the scan results for the critical and nucleation temperatures $T_c$ (left) and $T_n$ (right) of a FOEWPT, in the plane of the heaviest CP-even scalar mass $m_{h_3}$ vs.\ its singlet component $\Sigma_{h_3}=|R_{33}|^2$.  The lower-right region of the plots
is excluded due to the perturbative unitarity constraints (as discussed in section~\ref{SNR_FOEWPT}). 
For the gray points, the EW symmetry is not restored up to the maximum temperature used for this numerical analysis 
(recall the discussion in section~\ref{SNR_FOEWPT}), 
$T_{\rm max}=600\gev$,
and no FOEWPT takes place ($T_c$ and $T_n$ are
thus not defined).\footnote{One could
also argue that the gray points avoid the
problem of vacuum trapping, because no
FOEWPT has to take place in order to reach
the EW minimum at $T= 0$. Similar solutions
were proposed in the context of supersymmetric GUT
theories~\cite{Weinberg:1982id,Bajc:1999cn}.}
For the colored points in the left plot
of Figure~\ref{Jamaica_scan} there exists a critical temperature $T_c$ at which there are two degenerate minima, 
of the form $(0,0,\tilde{v}_S(T))$ and $(v_1(T),v_2(T),v_S(T))$, and one could naively be led to conclude that
these points also feature a FOEWPT. However, looking at the right plot of Figure~\ref{Jamaica_scan}, one can see that only a fraction of these points yield a FOEWPT occuring at $T = T_n$.
The black points in the right plot have $T_c$ defined,
but there is
no temperature $T < T_c$ for which the nucleation criterion of \refeq{eq:_TN} is fulfilled. As a consequence, the Universe stays trapped in the false vacuum $(0,0,\tilde{v}_S(T))$, and the EW phase transition does not occur.

In agreement with the results from~\cite{Basler:2019iuu}, we observe from Figure~\ref{Jamaica_scan} that for large values of the mass $m_{h_{3}}$ a FOEWPT is associated with a very singlet-like state $h_3$.
This is also reflected in the fact that for a fixed value of $m_{h_3}$ both $T_c$ and $T_n$ decrease with increasing $\Sigma_{h_3}$. 
On the other hand, for smaller values of $m_{h_3}$ a FOEWPT is possible for a sizeable doublet component in $h_3$. Moreover, for a fixed value of $\Sigma_{h_3}$, $T_c$ and $T_n$ decrease with decreasing $m_{h_3}$. 
This may be understood as follows:
Higgs bosons that participate in the EW phase transition (by acquiring a vev) should not be too heavy, since large Higgs boson masses require in general large bilinear terms, which hinder a FOEWPT if they enter the transition dynamics~\cite{Dorsch:2016nrg}.
In addition, the trilinear terms generating the potential barrier between true and false vacua are absent at tree-level in the N2HDM, and arise only from the radiative and thermal corrections to the potential, thus depending on the quartic scalar couplings $\lambda_i$. The values of $\lambda_i$ (and therefore the size of the potential barrier) grow with the splitting between the masses of $h_{2,3}$ and $A$ (where $m_A$ has been fixed at $650\gev$ in our scan). The strongest
FOEWPTs are then expected to occur for low values of ${m_{h_3}
\approx m_{h_2} \ll m_A}$, except when
$h_3$ is almost entirely singlet-like, 
i.e.~$\Sigma_{h_3} \approx 1$.

For a given $\Sigma_{h_3}$, there is a 
critical value of $m_{h_{3}}$ below which the energy barrier becomes so large that the probability for the tunnelling between vacua is too small to allow for the onset of the phase transition as defined in \refeq{eq:_TN}. The corresponding black points in Figure~\ref{Jamaica_scan} thus yield trapped metastable singlet vacua $(0,0,\tilde{v}_S)$ down to $T \to 0$. As discussed in section~\ref{sec:thermalhistory}, this situation yields an inflationary process that suffers from the ``graceful-exit'' problem~\cite{Guth:1982pn} and leads to an unphysical scenario. This is the case even though a critical temperature $T_c$ (at which the EW minimum becomes the global minimum of the potential) does exist for such points, as shown in the left plot of Figure~\ref{Jamaica_scan}. 
Furthermore, Figure~\ref{Jamaica_scan} highlights that this trapped-vacua region features the lowest values of $T_c$. In EW baryogenesis scenarios, the strength of the FOEWPT is precisely quantified as (see e.g.~\cite{Quiros:1999jp}) $\xi = v_c/T_c$ (where $v_c = \sqrt{v_1(T_c)^2 + v_2(T_c)^2}$ is the EW vev at the critical temperature). 
In an investigation based only on $T_c$ one would then naively -- and erroneously -- conclude that the strongest FOEWPTs for EW baryogenesis would occur in the region of parameter space corresponding to the black points in Figure~\ref{Jamaica_scan}. However, our results show 
that this region is unphysical. 

\medskip

Overall, the black ``vacuum-trapping'' region constitutes a sizeable fraction of the parameter space in our
scan that based on the thermal history of the Universe is ruled out because the breaking of the EW symmetry does not occur. This result makes manifest an important shortcoming of a zero temperature analysis of the stability of the EW minimum in extended scalar sectors (as implemented e.g. in \texttt{ScannerS}, see section~\ref{sec:vacstabzerotemp}), as we demonstrate that the presence of a global EW minimum at $T=0$ is not a sufficient criterion for an acceptable vacuum configuration in the N2HDM (this has also been shown recently within the NMSSM~\cite{Baum:2020vfl}).
If further local minima besides the EW minimum are present at $T = 0$, an analysis of the thermal history of the Universe including the nucleation probabilities of possible metastable minima is necessary to assess whether an N2HDM scenario is physical.
An analysis based only on the critical temperature (as done e.g.\ in~\cite{Basler:2019iuu}), is not sufficient and can yield misleading predictions.
We also note that an analysis of the N2HDM thermal history based on the Parwani resummation scheme would lead to a larger region of the parameter space experiencing vacuum-trapping: The Parwani method typically predicts smaller values for $T_c$ as compared to the Arnold-Espinosa method used in this work, since in the former the finite-$T$ contributions tend to restore the EW symmetry at lower temperatures (the same was found in the 2HDM~\cite{Basler:2016obg}).
The tunnelling probability scales with $\mathrm{exp}(- S_3 / T)$ (see \refeq{eq:_TN}), and so the onset of the EW phase transition is further suppressed in the Parwani resummation method.\footnote{We have found that all our 2000 scan points feature trapped-vacua when using the Parwani method, despite the EW minimum being the global minimum at $T=0$ for all of them.}
While we regard the Arnold-Espinosa method as more appropriate for 
the analyses in our paper (see the discussion above), the comparison with 
the Parwani resummation method shows that the parameter regions in the 
N2HDM that we have indicated as unphysical because of vacuum-trapping are 
robust and conservative, in the sense that those regions would be identified as unphysical based on both methods. 

\medskip

As a further step, we now discuss the 
interplay between the thermal evolution and the collider phenomenology of the N2HDM.
For the case of the 2HDM it has been found that the occurrence of a
FOEWPT is favoured by a hierarchical spectrum~\cite{Dorsch:2013wja,Basler:2016obg}, and
the decay $A \rightarrow Z h_{i}$ has emerged as a ``smoking-gun'' collider signature~\cite{Dorsch:2014qja} of a FOEWPT in the 2HDM.
Also in the N2HDM such a type of signature is linked with the possible presence of a FOEWPT, but the collider phenomenology related to this class of processes in the N2HDM is much richer than in the 2HDM.
In the alignment limit\footnote{We note that this limit is strongly
preferred for a FOEWPT in the 2HDM, see e.g.~\cite{Dorsch:2017nza}.} (which is realized 
in our parameter scan), the $A Z h_{\rm 125}$ coupling between the pseudoscalar $A$, the $Z$ boson  and the SM-like Higgs boson at about 125~GeV vanishes at tree-level. 
While in the 2HDM in this limit only the decay $A \rightarrow ZH$ is possible if kinematically allowed,
in the N2HDM the two decays $A \rightarrow Z h_{2}$ and
$A \rightarrow Z h_{3}$ can occur, whose branching ratios depend on
both the singlet component  
and the masses of $h_{2,3}$. As shown in Figure~\ref{Jamaica_scan}, these parameters also play an important role for the thermal history of the N2HDM. In our parameter scan we find that both decay channels are generally open in scenarios with a FOEWPT, except when $h_3$ is very singlet-like (and can thus effectively decouple from the FOEWPT dynamics, $m_{h_3} \gg v$).

\begin{figure}%[t]
\centering
\includegraphics[width=0.55\textwidth]{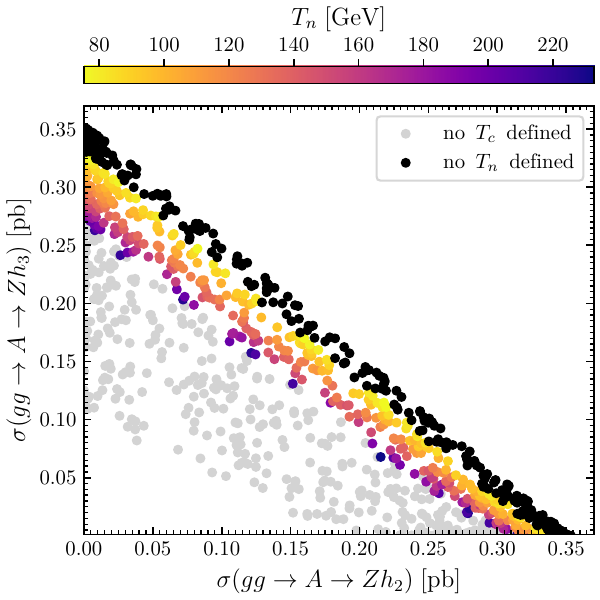}
\caption{\small Correlation of the cross sections for the processes
$A \rightarrow Z h_{2}$ and $A \rightarrow Z h_{3}$ for the N2HDM benchmark scenarios defined 
in Table~\ref{tab:input_parameters_Scan}.
The color coding is the same as
in Figure~\ref{Jamaica_scan} (right).
}
\label{BR}
\end{figure}

In Figure~\ref{BR} we show as result of our parameter scan defined in Table~\ref{tab:input_parameters_Scan} 
the predictions for the signal rates $pp\,(gg) \to A \rightarrow Z h_{2}$ and
$pp\,(gg) \to A \rightarrow Z h_{3}$
at the LHC with $\sqrt{s} =13$~TeV, where the production 
cross section has been calculated with
\texttt{SusHi v.1.6.1}~\cite{Harlander:2012pb,
Harlander:2016hcx},
and the branching ratios have been obtained
with \texttt{N2HDECAY}~\cite{Muhlleitner:2016mzt,Engeln:2018mbg}.
Since the production cross section ${\sigma(gg \rightarrow A)}$ is constant in our scan (it only depends on $m_A$ and $\tan\beta$), Figure~\ref{BR} effectively shows the interplay between BR$(A \rightarrow Z h_{3})$ and BR$(A \rightarrow Z h_{2})$. 
As a result, we find 
that (stronger) FOEWPTs with smaller nucleation temperatures are correlated with larger values for these branching fractions.
However, 
the largest values of the signal 
rates for each of the two processes in our scan correspond to unphysical trapped-vacua scenarios.
The detection of the processes $pp \to A \rightarrow Z h_{2}$ and 
$pp \to A \rightarrow Z h_{3}$ at the LHC would open the possibility 
to infer details about the thermal history of the Universe that would have occurred in the N2HDM.
Regarding the current status of LHC searches of this kind,
ATLAS and CMS have searched for the $pp \to A \rightarrow Z h_{i}$ (with $h_i \neq h_{125}$) signature within their $8\tev$~\cite{Khachatryan:2016are} and $13\tev$~\cite{Aaboud:2018eoy,Sirunyan:2019wrn} data sets, assuming that the Higgs boson $h_i$ decays into a pair of bottom quarks or a pair of $\tau$-leptons. It should be noted that our scan shows that for scenarios featuring a FOEWPT in the N2HDM the masses of both $h_2$ and $h_3$ could easily be above the decay threshold into top-quark pairs. 
In fact, for the rather small value of $\tan\beta=2$ in our scan the
discovery potential for the
``smoking-gun'' signatures in the N2HDM scenarios could be higher for the decay of $h_{2,3} \to \bar{t}t$. Thus, our results motivate to explore the
signature ${pp \to A \rightarrow Z (h_{i}) \to Z (\bar{t}t)}$
within the programme of experimental searches at the LHC (see
also~\cite{Haisch:2018djm}).

\subsection{Case 2: Singlet admixture in \texorpdfstring{$h_{125}$}{hSM}}
\label{mixhsmsing}

In contrast to the scan studied in the previous section,
we now explore scenarios where the Higgs boson $h_{125}$ has a singlet admixture (whereas the heavy CP-even state $H$ does not), and study the impact of such an admixture on the N2HDM thermal history.
The measurements of the signal rates of the SM-like Higgs boson together with the EWPO set limits on the possible amount of a singlet component that can be acquired by $h_{125}$~\cite{Muhlleitner:2016mzt,Biekotter:2019kde}. These limits also constrain the possible impact of the singlet-doublet mixing on the FOEWPT in the considered scenario.
In order to be able to study the effect of a singlet admixture in $h_{125}$ over a substantial mixing 
range, here we will fix the mass of the singlet-like scalar in our parameter scan to be relatively close to 125 GeV (the relatively small mass splitting between the Higgses then allows for sizeable mixing).
We perform two N2HDM parameter scans with \texttt{ScannerS}, defined in Table~\ref{tab:input_parameters_hsmlightes}, to cover both mass orderings (as they have different phenomenological implications):
a singlet-like scalar 
somewhat heavier than $h_{125}$, 
and a singlet-like scalar 
somewhat lighter than $h_{125}$.
In both scans
we keep the hierarchy between the masses of the heavier CP-even
doublet-like Higgs boson ($h_b$ in Table~\ref{tab:input_parameters_hsmlightes}) and the CP-odd state $A$ in order to guarantee the presence of a potential barrier separating the false minimum at $(0,0,\tilde{v}_S)$ and the true minimum at $(v_1,v_2,v_S)$.

\begin{table}
\centering
{\renewcommand{\arraystretch}{1.4}
\addtolength{\tabcolsep}{-1pt}
\small
\begin{tabular}{cccccccccccc}
 $m_{h_{a}}$  & $m_{h_{b}}$  & $m_{h_{c}}$ & $m_{A}$ &
$m_{H^{\pm}}$  & $\text{tan}\beta$  & $C^{2}_{h_{a}t\bar{t}}$ & $C^{2}_{h_{a}VV}$
&
$\mathrm{sgn}(R_{a3})$  & $R_{b3}$  & $m_{12}^{2}$ & $v_{S}$\\
\hline 
\hline 
$125.09$  & $400$  & $160$ & $650$ &
$650$  & $2$  & $[0.8,1.2]$ & $[0.7,1.0]$ &
$1$  & $0$  & $65000$ & $300$ \\
\hline
$125.09$ & $400$  & $105$ & $650$ &
$650$  & $2$  & $[0.8,1.2]$ & $[0.7,1.0]$ &
$-1,1$  & $0$  & $68500$ & $300$
\end{tabular}
\addtolength{\tabcolsep}{1pt}
}
\caption{\small \texttt{ScannerS} input parameters used in the study of the impact of a singlet admixture in  $h_{125}$ on the N2HDM thermal history. The upper (lower) row corresponds to the scan parameters for $m_{h_c} > (<) \,125.09 \gev$.}
\label{tab:input_parameters_hsmlightes}
\end{table}

\medskip

\begin{figure}
\centering
\includegraphics[width=0.44\textwidth]{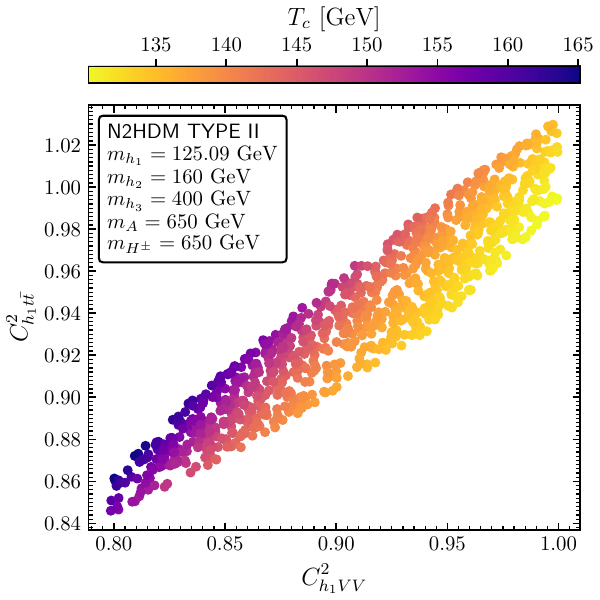}
\hspace{0.02\textwidth}
\includegraphics[width=0.44\textwidth]{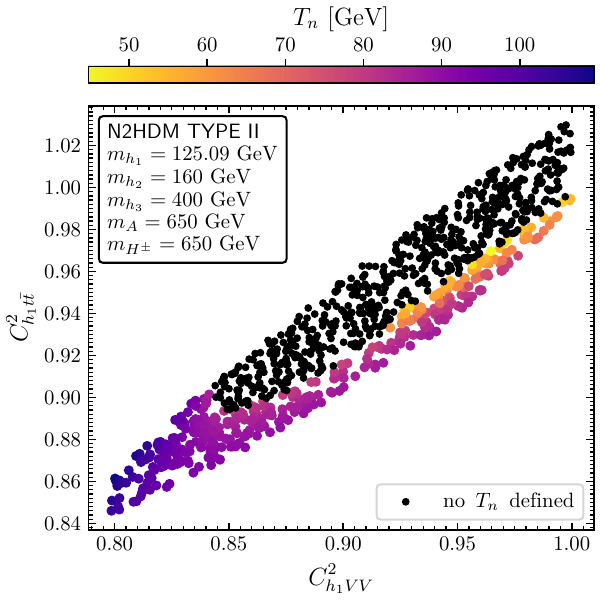}
\caption{\small Parameter scan according to Table~\ref{tab:input_parameters_hsmlightes} (upper row) in the $C_{h_1 VV}^2$--$C_{h_1 t \bar t}^2$ plane.
The color coding indicates the values of the 
critical temperature $T_c$ (left)
and the nucleation temperature $T_n$ for the points with a FOEWPT (right).
For the black points the Universe is trapped in a false minimum, 
such that no
nucleation temperature can be defined.
}
\label{fighSMlight}
\end{figure}

We first analyse the scenario in which $h_{125}$ is the lightest Higgs boson $h_1$, choosing our \texttt{ScannerS} scan parameters as shown in the first row of Table~\ref{tab:input_parameters_hsmlightes}. We generate 1000 benchmark points fulfilling the theoretical and experimental constraints discussed in section~\ref{constraints}. The only parameters that are varied in our scan are the scalar mixing angles, under the condition that the singlet component of $h_b = h_3$ vanishes. We choose the mass of the singlet-like Higgs boson $m_{h_2}=160\gev$ to allow for a sizeable mixing with the SM-like Higgs boson $h_1$ without being in conflict with the LHC searches for scalar resonances decaying into a pair of $Z$ bosons~\cite{Sirunyan:2018qlb}. 
In Figure~\ref{fighSMlight} we show the set of parameter points in the plane of normalized squared-couplings of the Higgs boson at 125~GeV to EW gauge bosons and SM fermions, $C_{h_1 VV}^2$ and $C_{h_1 t \bar t}^2$, respectively.
Note that $C_{h_1 VV}^2$ also corresponds to the singlet component of $h_2$, i.e. $\Sigma_{h_2} = 1 - \Sigma_{h_1} = C_{h_1 VV}^2$ ($\Sigma_{h_3} = 0$ because of $R_{b3} = 0$).
The allowed parameter space in Figure~\ref{fighSMlight} is defined by three different constraints: boundedness from below of the tree-level potential requires $\lambda_1 > 0$, which excludes the lower-right triangular region of the ($C_{h_1 VV}^2$, $C_{h_1 t \bar t}^2$) plane in Figure~\ref{fighSMlight}; it also requires $\lambda_2 >0$, which excludes the upper-left triangular region of the ($C_{h_1 VV}^2$, $C_{h_1 t \bar t}^2$) plane in Figure~\ref{fighSMlight}. 
We then find a diagonal band of allowed parameter space, based only on these theoretical considerations.  Finally, the roughly elliptical shape of this allowed band in Figure~\ref{fighSMlight} is due to the experimental LHC constraints on the $h_{125}$ signal strengths, which we implement using the $\chi^2$ result of \texttt{HiggsSignals} (see section~\ref{sechSMconst}).

For each of the 1000 scan points, we have performed a finite-$T$ analysis with \texttt{CosmoTransitions}, 
computing the thermal evolution of the effective potential from a maximum temperature $T_{\rm max}=300\gev$\footnote{The maximum temperature $T_{\rm max}$ that we consider here is substantially lower than in our previous scans, since here we are not interested in the symmetry non-restoration behavior, but rather in the appearance of trapped vacua, for which the temperatures studied need not be much larger than the EW scale.}    
down to $T = 0$.
In contrast to the scan results discussed in section~\ref{jamaica}, here we find that 
at $T_{\rm max}$ the Universe finds itself in a minimum of the kind $(0,0,\tilde{v}_S)$ for all N2HDM benchmarks, and so the EW symmetry is always restored (at least in an intermediate temperature regime).
As the Universe cools down from $T_{\rm max}$, all benchmark points feature a critical temperature $T_c$, shown in Figure~\ref{fighSMlight} (left), at which the EW-broken minimum is degenerate with the singlet minimum $(0,0,\tilde{v}_S)$. For 609 points in our scan 
the Universe remains trapped in the false (singlet) vacuum, as indicated by the black points in the right plot of Figure~\ref{fighSMlight}, while for the remaining (391) points a FOEWPT takes place. In the latter case, the nucleation temperature $T_n$ is shown in Figure~\ref{fighSMlight} (right).
We see that both $T_c$ and $T_n$ decrease with increasing $C_{h_1VV}^2$, in agreement with
results previously obtained in the 2HDM~\cite{Bernon:2017jgv,Basler:2016obg}. This suggests (as we have discussed also in the previous sections) that the strength of the transition reaches the largest values in the alignment limit, in which the heavier doublet-like CP-even
Higgs boson does not obtain a vev.
We note that within the 2HDM this reduces the prospects for 
detecting deviations from the SM case via signal rate  measurements of $h_{125}$ in the parameter space that is relevant for FOEWPTs. Our results show that this is also the case for the N2HDM if $h_{125}$ is the lightest Higgs boson. However, the opposite effect can occur 
if $h_{125}$ is the second-lightest Higgs boson,
as we will show below.
 
\bigskip

\begin{figure}
\centering
\includegraphics[width=0.44\textwidth]{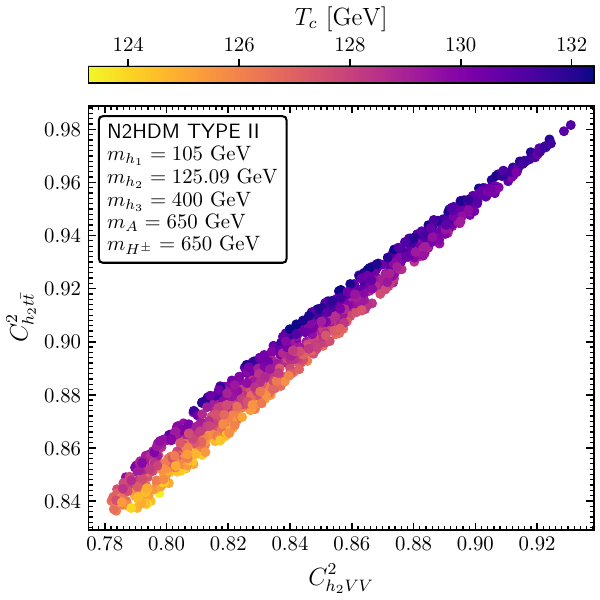}
\hspace{0.02\textwidth}
\includegraphics[width=0.44\textwidth]{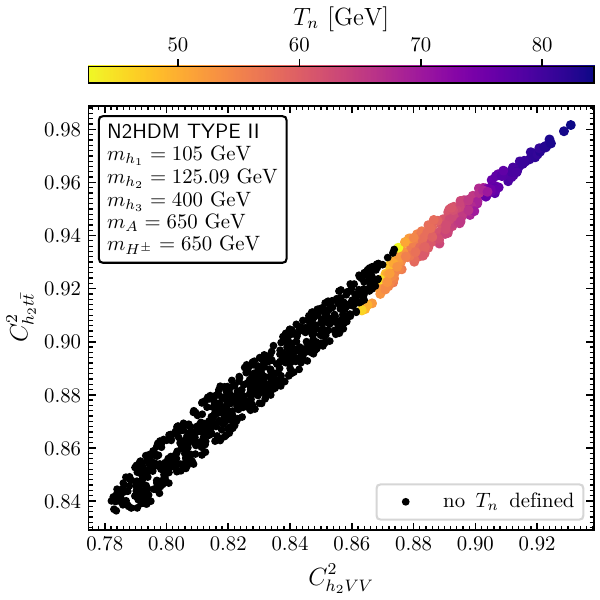}
\caption{\small Parameter scan according to Table~\ref{tab:input_parameters_hsmlightes} (lower row) in the $C_{h_2 VV}^2$--$C_{h_2 t \bar t}^2$ plane.
The color coding indicates the values of the 
critical temperature $T_c$ (left)
and the nucleation temperature $T_n$ for the points with a FOEWPT (right).
For the black points the Universe is trapped in a false minimum, 
such that no
nucleation temperature can be defined.
}
\label{fighSMnext}
\end{figure}

We turn now to the analysis of scenarios where the singlet-like scalar state is lighter than $h_{125}$, 
in order to demonstrate
the importance of the mass ordering of the singlet- and doublet-like Higgs bosons. Our parameter scan now corresponds to the lower row of
Table~\ref{tab:input_parameters_hsmlightes}, with
${m_{h_1} = 105\gev}$. This
allows for a sizeable variation of the mixing with $h_{125} = h_2$ without being in conflict
with the cross section limits obtained from the  LEP Higgs searches~\cite{Barate:2003sz}.
We also slightly increase the value of $m_{12}^2$ compared to the previous analysis (see the upper row of Table~\ref{tab:input_parameters_hsmlightes}) in order to increase
the tunnelling rate between minima in a FOEWPT.\footnote{As shown in~\cite{Bernon:2017jgv} for the 2HDM, larger values of $m_{12}^2$ reduce the potential barrier and the distance in field space between false
and true minima, thus increasing the tunnelling rate of FOEWPTs.
In the present scan, the increased value of $m_{12}^2$
counterbalances the otherwise suppressed tunnelling
probability due to the overall reduced mass scale of the
CP even Higgs bosons when $m_{h_1} = 105\gev$ and $m_{h_2} = 125\gev$
compared to the previous scan with $m_{h_1} = 125\gev$ and
$m_{h_2} = 160\gev$.}
We generate 1000 points fulfilling the theoretical and experimental constraints
using \texttt{ScannerS} and use \texttt{CosmoTransitions} to analyse the thermal history of each scan point as described above for the previous scan.
In Figure~\ref{fighSMnext} we show the resulting points in the plane of 
the effective couplings $C_{h_2 VV}^2$ and $C_{h_2 t \bar t}^2$. The allowed parameter space is defined as in Figure~\ref{fighSMlight} by the bounded-from-below constraints $\lambda_{1,2} > 0$ and by the constraints on the signal rates of $h_{125}$ that are tested with~\texttt{HiggsSignals}. 
It is interesting to note that the region with 
$C_{h_2 VV}=C_{h_2 t \bar t}=C_{h_2 b \bar b}=1$,
i.e.\ the alignment limit, cannot be realized with the choice of parameters of this scan, since in this region one finds $\lambda_1 < 0$. 
For the allowed points, we find $0.5 \lsim C_{h_2 b \bar b} \lsim 0.8$
(not shown in the plots) 
together with $C_{h_2 VV} < 1$ and $C_{h_2 t \bar t} < 1$
in order to satisfy the constraints on the signal rates
of $h_{125}$.\footnote{For $C_{h_{125} VV}^2,
C_{h_{125} t \bar t}^2 < 1$ 
and $C_{h_{125} b \bar b}^2 \approx 1$ the
diphoton branching ratio of $h_{125}$ would
be too small.}

All 1000 points in our scan feature a critical temperature $T_c$ (at which the true EW minimum is degenerate with the false minimum $(0,0,\tilde{v}_S)$), shown in the left plot of Figure~\ref{fighSMnext}. 
However, the majority of scan points correspond to trapped-vacuum scenarios, shown in the right plot of Figure~\ref{fighSMnext} as black points, and excluded since a FOEWPT does not take place (from the 1000 points only 225 yield a FOEWPT).
For the points that do feature a FOEWPT, the color coding in Figure~\ref{fighSMnext} (right) indicates the nucleation temperature $T_n$. 
As opposed to the scenario with $h_{125}$ as the lightest Higgs boson, here
both $T_c$ and $T_n$ are reduced for decreasing values of $C_{h_{125} V V}$.
A decrease in $ C_{h_{125} VV}$ is linked to an increase in the mixing between the (singlet-like) lightest Higss boson $h_1$ and $h_{125}$, $\Sigma_{h_{125}} \approx 1-C_{h_{125} VV}^2$, which decreases the mass scale of the particles obtaining a vev during the transition. 
This in turn increases the strength of the FOEWPT and leads to a lower $T_n$.
As Figure~\ref{fighSMnext} highlights, this decrease in $T_n$ eventually results in vacuum-trapping, which in our scan rules out all points with $\Sigma_{h_{125}} \gtrsim
0.15$.
The fact that, given the presence of a singlet-like scalar below $125\gev$, the N2HDM can realize a FOEWPT quite far away from the alignment limit is a very important difference w.r.t.~the 2HDM (in which the strength of the FOEWPT is maximized in the alignment limit~\cite{Basler:2016obg,Dorsch:2017nza}).
In the N2HDM, a mixing of $h_{125}$ with a lighter singlet-like scalar reverses the dependence of the EW phase transition strength  on $C_{h_{125} VV}$ compared to the case of the 2HDM. This constitutes a key feature of the N2HDM regarding the interpretation of possible deviations from the SM in the 
couplings of the Higgs boson at 125~GeV that could be revealed in future measurememts at the LHC and the HL-LHC.

\section{Conclusions}
\label{conclu}

Extensions of the SM Higgs sector allow for a rich cosmological history associated to the thermal evolution of the scalar potential in the early Universe. In this work we have focused on the N2HDM,
an extension of the SM in which the Higgs sector is supplemented by a second Higgs doublet and a real scalar singlet field. 
We have found that within the N2HDM the evolution of the early Universe
can give rise to a very rich phenomenology, so far poorly explored for other extended Higgs sectors.
Besides the possibility of a FOEWPT, which has been studied in depth for all these scenarios, 
we have shown that within the N2HDM further phenomena can occur, namely
the non-restoration of the EW symmetry at high temperatures and the existence of false vacua in which the Universe gets trapped at $T \to 0$ (which we refer to as \textit{vacuum trapping}). We have demonstrated that both types of scenarios have important consequences for the phenomenological viability of N2HDM scenarios, and expect them to be relevant also for wider classes of models.

We have first studied the phenomenon of EW symmetry non-restoration, which has recently regained attention in the literature~\cite{Meade:2018saz,Glioti:2018roy,Baldes:2018nel,Bai:2021hfb}. We have shown that in the N2HDM this behaviour is  driven by contributions from the resummation of daisy diagrams.
We have identified the key quantities that can be used to analytically determine the restoration or non-restoration of the EW symmetry at high temperature, summarized in Eqs.~(\ref{coeff_1})--(\ref{coeff_3}) and \eqref{H11S_limit}.
Even though we focused on the type~II N2HDM,
our analytical approach can also be applied
to the other Yukawa types of the model,
as it mainly depends on the quartic
scalar couplings and the Yukawa coupling of
the top quark.
We have supplemented our analytical investigation with a numerical analysis of the N2HDM thermal history  
with the help of the code \texttt{CosmoTransitions}, tracking in each case the local minima of the potential as a function of temperature. We have also studied the relation of EW symmetry non-restoration at high $T$ to the dynamics of the Higgs potential at temperatures close to the EW scale, and showed that it is possible for a scenario with an unrestored
EW symmetry at high temperatures to still feature a FOEWPT, since the EW symmetry can be restored in an intermediate temperature regime. 

As a further step, we have discussed the occurrence and the impact of vacuum trapping in the N2HDM. In contrast to previous studies of the N2HDM, which only relied on the existence of a critical temperature at which the EW phase becomes the deepest potential minimum, we have demonstrated that 
an investigation of the transition probability 
to the EW minimum in the early Universe is crucial to assess the physical viability of the N2HDM parameter space. In particular, we have shown that scenarios featuring trapped vacua  
lead to unphysical regions of the N2HDM parameter space despite the presence of a global EW minimum of the scalar potential at $T = 0$. 
As a consequence, our analysis reveals that sizeable parts of the otherwise unconstrained N2HDM parameter space are ruled out because they would give rise to vacuum trapping. 
In addition, we have studied the interplay between vacuum trapping, 
EW symmetry non-restoration and the possibility of a FOEWPT.

Finally, we have analyzed the connection of these early Universe phenomena to the predicted phenomenology of the N2HDM at the LHC.  We have shown that the patterns of the thermal history of the early Universe can be linked to characteristic signatures in the N2HDM which have no equivalent in other models like the 2HDM, in particular possible signals in the two decay channels $A \to Z h_2$ and $A \to Z h_3$. We have also shown that in the N2HDM a departure from the alignment limit does not necessarily diminish the prospects for a FOEWPT, in contrast to the case of the 2HDM. 
Our results suggest that the combination of constraints from 
collider experiments, from the evolution of the early Universe and from future 
astrophysical experiments, such as gravitational wave interferometers, will be very valuable for probing the parameter space of the N2HDM and of further BSM models featuring an extended Higgs sector.

\section*{Acknowledgements}
We thank
D.~Curtin,
Y.~Gouttenoire, 
T.~Konstandin,
Z.~Liu,
P.~Meade,
Y.~Wang
and
J.~Wittbrodt
for useful discussions. The work of T.B., M.O.O.\ and G.W.\ is supported by the Deutsche
Forschungsgemeinschaft under Germany’s Excellence Strategy EXC2121 ``Quantum
Universe'' - 390833306.
The work of J.M.N.\ is supported by the Ram\'on y Cajal Fellowship contract RYC-2017-22986, and by grant PGC2018-096646-A-I00 from the Spanish Proyectos de I+D de Generaci\'on de Conocimiento.
J.M.N.\ also acknowledges support from the European Union's Horizon 2020 research and innovation programme under the Marie Sklodowska-Curie grant agreement 860881 (ITN HIDDeN). The work of S.H.\ is supported in part by the
MEINCOP Spain under contract PID2019-110058GB-C21. S.H. and J.M.N.\ are also supported by 
the AEI through the grant IFT Centro de Excelencia Severo Ochoa SEV-2016-0597.

\bibliography{draft.bib}
\bibliographystyle{kp}

\end{document}